\documentclass[12pt]{article}      
\usepackage{graphicx}

\usepackage{amsmath,amsfonts,amssymb}
\newtheorem{theorem}{Theorem}

\newtheorem{proposition}{Proposition}
\newtheorem{corollary}{Corollary}
\newtheorem{lemma}{Lemma}
\newtheorem{hypothesis}{Hypothesis}

\begin{document}

\title{Fixation probabilities for the Moran process in evolutionary games with two strategies: graph shapes and large population asymptotics
\thanks{EPS had scholarships paid by CNPq (Conselho Nacional de Desenvolvimento Cient\'ifico e Tecnol\'ogico, Brazil) and Capes (Coordena\c c\~ao de Aperfei\c coamento de Pessoal de N\'ivel Superior, Brazil). EMF had a Capes scholarship. AGMN was partially supported by Funda\c{c}\~ao de Amparo \`a Pesquisa de Minas Gerais (FAPEMIG, Brazil).}
}
\author{Evandro P. Souza \and Eliza M. Ferreira \and Armando G. M. Neves
}

\author
{Evandro P. Souza$^{1}$ \\ Eliza M. Ferreira$^{2}$ \\Armando G. M. Neves$^{1}$
	\\
	\normalsize{$^{1}$Departamento de Matem\'atica, Universidade Federal de Minas Gerais,}\\
	\normalsize{$^{2}$Departamento de Ci\^encias Exatas, Universidade Federal de Lavras}\\
	\normalsize{aneves@mat.ufmg.br}
}

%\titlerunning{Graph shapes and large populations}        % if too long for running head

\date{April 30, 2018}

\maketitle

\begin{abstract}
This paper is based on the complete classification of evolutionary scenarios for the Moran process with two strategies given by Taylor et al. (B. Math. Biol. 66(6): 1621--1644, 2004). Their classification is based on whether each strategy is a Nash equilibrium and whether the fixation probability for a single individual of each strategy is larger or smaller than its value for neutral evolution. We improve on this analysis by showing that each evolutionary scenario is characterized by a definite graph shape for the fixation probability function. A second class of results deals with the behavior of the fixation probability when the population size tends to infinity. We develop asymptotic formulae that approximate the fixation probability in this limit and conclude that some of the evolutionary scenarios cannot exist when the population size is large.
\end{abstract}

\section{Introduction}
\label{secintro}
The Moran process \cite{moran} was introduced as a stochastic model for the genetic evolution of a haploid population with asexual reproduction, assuming no mutations, and fixed finite size. An important feature of the Moran process -- and other similar stochastic evolutionary processes for finite populations -- is \textit{fixation}: with probability 1, at a finite time the whole population will be made of a single type of individuals. This is a consequence of the finiteness of the population and of the no mutations assumption.

The subject of fixation in evolutionary processes in finite populations, using either the Moran process or others, has received a good deal of attention in recent publications, see e.g.\cite{ChalubSouza2009}, \cite{ChalubSouza2016}, \cite{XuZhangetal}, \cite{AshcroftAltrockGalla} and \cite{DurandLessard}. In all of the above and most others, the population is composed of only two types of individuals, say type A and type B, where A and B may stand for particular alleles at some genetic locus, or phenotypes. Some papers dealing with cases of populations with more than two types are \cite{Mobilia},\cite{Durneyetal} and \cite{TraulsenClaussenHauert}.

In this paper we will restrict only to the case in which the population is composed of just two types of individuals. We will suppose the population has a fixed sized and use symbol $N$ for the number of individuals in it. For the Moran process for a population with two types it is possible to obtain exact results for some quantities of interest \cite{ewens}. One such quantity is the type A fixation probability $\pi_i$, i.e. the probability that the population becomes all A, given that the initial state is $i$ A individuals and $N-i$ B individuals, $i\in \{0, 1, 2, \dots, N\}$. Such an exact result is not available either for the Moran process if the number of types of individuals in the population is larger than two, or for other processes in Population Genetics, such as Wright-Fisher.

The Moran process was originally introduced for the case in which the fitness of the individuals in the population depends on the individual's type (A or B), but not on the frequency of that type in the population. In \cite{nowaknature}, \cite{taylor} it was extended to the case of Evolutionary Game Theory \cite{MaynardSmithPrice}, \cite{hofbauersigmund}, \cite{nowakbook}, in which  fitnesses depend on the frequency in which the types are present in the population. In the context of Game Theory, the types of individuals are commonly referred to as strategies.

At first, Evolutionary Game Theory was intended as a theory for infinite populations with a deterministic dynamics \cite{taylorjonker}, \cite{hofbauersigmund}. An important application of this approach is the study of models for the evolution of cooperation \cite{nowaksigmundnature}, \cite{nunezneves}. For the deterministic replicator dynamics and populations with two strategies, there exist 4 possible generic (or robust, see \cite{zeeman}) dynamical behaviors: dominance of A, dominance of B, coordination and hawk-dove. The above classification depends on whether the pure strategies A and B are Nash equilibria.

The Moran process, on the other hand, allows for the possibility of random fluctuations, as expected for finite populations. It happens that the result of the evolutionary process is not always the fixation of the fittest type. \cite{taylor} elucidated in a clear way the important role played by the population size $N$ in the outcome of evolution in the game theoretical Moran process. They also noticed that taking into account the random nature of the Moran process led to a further complexity step necessary to classify the dynamics of a population.Their classification scheme takes into account not only whether the pure strategies A and B are Nash equilibria. It considers also whether the fixation probabilities $\rho_A \equiv \pi_1$ of a single A individual in a population of $N-1$ B individuals and its analogue $\rho_B\equiv 1-\pi_{N-1}$ are larger or smaller than $1/N$. Here $1/N$ appears as the fixation probability for the neutral Moran process \cite{nowakbook}, see Section \ref{secbackground} for precise definitions. In principle, the possible number of evolutionary scenarios for the Moran process in a population with two types of individuals is 16. An important result in \cite{taylor} is that only 8 out of the 16 scenarios are allowed to exist. We stress that the classification scheme of \cite{taylor} takes into account $\rho_A$ and $\rho_B$, i.e. the values of the fixation probabilities $\pi_i$ for only two values of $i$, namely $i=1$ and $i=N-1$.

As already remarked, an explicit exact formula for $\pi_i$ exists, see (\ref{piiformula}). Despite that, the formula is ``unwieldy", as remarked in \cite{ewens}, and this unwieldiness is the reason for this paper. One result we will show is that given the evolutionary scenario as classified by \cite{taylor}, we can deduce properties of $\pi_i$ for the remaining values $i=2, 3, \dots, N-2$. More than that, we can associate with each evolutionary scenario a precise \textit{shape} for the graph of the $\pi_i$ as a function of $i$. We will see e.g. that the $\pi_i$ graph may either lie entirely above the neutral line $\pi_i=i/N$, or entirely below it, or is allowed to intercept it only once, each of the above possibilities being associated with some of the possible scenarios. We will also show that a single ``inflection point" will exist for some scenarios and no inflections will be allowed for some other scenarios. In this analysis, instead of using the exact formula for the $\pi_i$, we will use the discrete derivative of $\pi_i$, which obeys a much simpler equation.

A second class of results in the present paper deals with another gap in the analysis of \cite{taylor}. In fact, reading the above paper, we felt that some analysis was missing on the relation between what happens for large $N$ and the corresponding deterministic $N=\infty$ case. We will prove e.g. that for large population sizes $N$ only 5 out of the 8 evolutionary scenarios are generically admissible. More precisely, we will show that if we fix the pay-off matrix of the game and the intensity of selection $w$, see Section \ref{secbackground} for precise definitions, then for large enough $N$ only some of the evolutionary scenarios are allowed.  

The tools we developed for this second part of the paper are asymptotic formulas for $\pi_i$ when the fraction $i/N$ of A individuals is held fixed and $N \rightarrow \infty$. As we shall see, the starting point of the argument is identifying a certain sum as a Riemann sum and approximating it when $N \rightarrow \infty$ with the corresponding integral. It seems to us that this idea was discovered by \cite{AntalScheuring}, but they used it only as a non-rigorous approximation. Despite that, they are able to obtain qualitatively correct asymptotic results for the fixation probabilities $\rho_A$ and $\rho_B$ of a single mutant. Sometimes their approximation can lead to quantitatively correct results, too. They also study the question of fixation times, an issue not at all addressed in this paper.

\cite{ChalubSouza2016} presented a rigorous theory in which the just mentioned integral is called the ``fitness potential". They define ``suitable birth and death processes", a class which contains the Moran process, and prove for these processes an interesting theorem in which the fixation probabilities are asymptotically calculated as $N \rightarrow \infty$ in terms of the fitness potential. In the class of suitable processes there is a wealth of interesting cases where the intensity of selection depends on $N$, but unfortunately the case which interests us, i.e. the intensity of selection $w$ independent of $N$, does not match their suitability hypotheses.

We believe that one reason why both \cite{AntalScheuring} and \cite{ChalubSouza2016} did not obtain all of our present results is that replacing the integral by a Riemann sum produces a ``continuation" error term that does not tend to 0 as $N \rightarrow \infty$. We obtain asymptotic formulae both for the main contribution to the fixation probability, i.e. the one individuated by \cite{AntalScheuring}, and also for the continuation error. As a consequence, we will see that some of the evolutionary scenarios are forbidden for large enough $N$. 

This paper is organized as follows. In Section \ref{secbackground} we will review the Moran process as defined in \cite{taylor}, but adding the possibility of a weak-selection regime, not accounted for in that work, although present in the contemporary paper \cite{nowaknature} by the same authors. We will also explain the notations and derive again the 8 possible evolutionary scenarios. The graph shapes for the fixation probability functions in each scenario are derived in Section \ref{secgraphshapes}. The general formulation of the asymptotics for large $N$ problem is presented in Section \ref{secasymptotics}. Sections \ref{seccoord} and \ref{secremaining} deal with the asymptotic results for the particular scenarios. In Section \ref{secconc} we draw some conclusions and suggest some lines to be followed in future works. Some of the lengthier proofs are found in Appendix \ref{appproofs}.

Before anything else, we introduce for definiteness some not so much standard notation to be used in this paper. First of all, when we talk about asymptoticity, it will always be in the infinite population limit $N \rightarrow \infty$, where $N$ is the number of individuals. We say that two functions $f$ and $g$ are asymptotic, denoted $f \sim g$, if  $\lim_{N \rightarrow \infty}f(N)/g(N)=1$. We will write $f=o(g)$ if $f(N)/g(N) \rightarrow 0$ when $N \rightarrow \infty$. If $f(N)/g(N)$ is bounded when $N \rightarrow \infty$, we write $f=O(g)$. 

We define also $\mathbb{Q}_N= \{\frac{i}{N} ; i \in \mathbb{Z}\}$. If $x \in \mathbb{R}$, then $[N x]$ denotes the integer closest to $Nx$. Of course, for $x \in \mathbb{Q}_N$, $[Nx]=Nx$.

\section{Background on the Moran process and notation}
\label{secbackground}
We will use the same notation as in \cite{nowaknature} for the Moran process and fitnesses. Let $N$ be the number of individuals in a population. Each individual may be either of type (or strategy) A or type B. For the sake of numbering rows and columns of the pay-off matrix, types A and B will be indexed as types 1 and 2, respectively.

The game-theoretic pay-off matrix is 
\begin{equation}  \label{payoffmatrix}
P \,=\,\left(\begin{array}{cc}
a&b\\
c&d
\end{array}
\right) \;,
\end{equation}
with $a$, $b$, $c$ and $d$ all positive. The matrix element in row $k$ and column $\ell$ is the pay-off for an individual of type $k$ interacting with an individual of type $\ell$. Considering the mean-field hypothesis that individuals interact with all others, but not with themselves, the game-theoretic pay-off for A individuals in a situation in which there are $i$ individuals of type A and $N-i$ individuals of type B in the population is
\[a \, \frac{i-1}{N-1} \,+\, b\, \frac{N-i}{N-1} \,.\]
Here $\frac{i-1}{N-1}$ is the fraction of A individuals other than the one receiving the pay-off in the remainder of the population and $\frac{N-i}{N-1}$ is the corresponding fraction of B individuals. By an analogous argument the game-theoretic pay-off for B individuals is
\[c \, \frac{i}{N-1} \,+\, d\, \frac{N-i-1}{N-1} \,.\]

In Evolutionary Game Theory, the game theoretic pay-off is considered to be part of individuals' fitness, i.e. a contribution to the size of their offspring. Other than the game-theoretic pay-off, we consider as in \cite{nowaknature} that the fitness of each individual has another term which is equal for all individuals in the population. Summing both terms, when the population has $i$ A and $N-i$ B individuals we define the  fitness of the A's as
\begin{equation}
\label{fi}
f_i \,=\, 1-w \,+\, w \, [a \, \frac{i-1}{N-1} \,+\, b\, \frac{N-i}{N-1}] \;
\end{equation}
and the fitness of B's as
\begin{equation}
\label{gi}
g_i \,=\, 1-w \,+\, w \, [c \, \frac{i}{N-1} \,+\, d\, \frac{N-i-1}{N-1}] \;.
\end{equation}
Parameter $w \in [0,1]$ in the last two equations will be referred to as \textit{intensity of selection}. For $w=1$, we are considering that the game-theoretic pay-off accounts for the totality of each individual's fitness, situation considered e.g. in \cite{taylor} and \cite{AntalScheuring}. For $w=0$, on the contrary, all individuals have the same fitness, independent of their type. This important particular case is called \textit{neutral}. When $w$ is small, we say we are in the \textit{weak-selection regime}.

In this paper we will consider the following
\begin{hypothesis}  \label{intensityhypothesis}
	The intensity of selection $w$ appearing in (\ref{fi}) and (\ref{gi}) does not depend on the population size $N$.
\end{hypothesis}
In \cite{ChalubSouza2009}  the above hypothesis is explicitly violated and some interesting results are obtained in the weak-selection regime by exactly choosing the form how $w$ depends on $N$. In a further paper \cite{ChalubSouza2016} the same authors go beyond weak-selection and present some asymptotic results when $N \rightarrow \infty$ using ideas closely related to the ones in this paper. But their main theorem does not apply to the case of $w$ independent of $N$ as here.

Returning to (\ref{fi}) and (\ref{gi}), notice that $f_i$ and $g_i$ in general depend on $i$, i.e. on the frequency of A individuals in the population, as usual in game-theoretic situations. The special case of frequency independent fitnesses, typical of Population Genetics, may be recovered taking  $a=b$ and $c=d$ in (\ref{payoffmatrix}).

Population dynamics is introduced when we allow the frequencies of A and B to evolve with time. In the Moran process \cite{moran}, this is modeled by two independent random choices at each time step: one random individual is chosen for reproduction and one random individual is chosen for death. We assume that no mutations occur in reproduction, i.e. that a new individual with the same type as the reproducing one replaces the dying individual. The number $N$ of individuals in the population remains constant for all time. The dying individual is chosen uniformly among the whole population, but the reproducing individual is chosen with probability proportional to the fitness of its type. 

With the above defined stochastic dynamics, at each time step the number $i$ of A individuals may increase by 1 (if an A is chosen for reproduction and a B for death), decrease by 1 (if a B is chosen for reproduction and an A for death), or remain constant (in the remaining two possible outcomes). The corresponding transition probabilities, taking into account the reproduction probabilities proportional to fitnesses, are
\begin{equation}  \label{alphai}
P_{i \rightarrow i+1} \, \equiv \, \alpha_i \, =\, \frac{i \, f_i}{i \, f_i+ (N-i) \, g_i} \, \frac{N-i}{N} \;,
\end{equation}
\begin{equation}  \label{betai}
P_{i \rightarrow i-1}\, \equiv \, \beta_i \,=\, \frac{(N-i) \, g_i}{i \, f_i+ (N-i) \, g_i} \, \frac{i}{N} \;,
\end{equation}
and 
\[P_{i \rightarrow i} \,=\, 1- \alpha_i - \beta_i\;.\]

If we take the number $i$ of A individuals, $i \in \{0,1, \dots, N\}$, as the state of the population, then the state evolves as a Markov chain \cite{allen}. States $i=0$ and $i=N$ are absorbing, i.e. the chain will stay forever in such a state if the chain ever enters it. It can be seen that all the other states $i=1, 2, \dots, N-1$ are transient \cite{allen}. Fixation, i.e. the chain eventually entering one of the absorbing states, will happen with probability 1. This is a consequence of Markov chain theory when the number of states is finite, see e.g. \cite{allen}.

Let $\pi_i$ be the probability that the process will end up in the state $N$ (i.e. fixation of type A) given that the initial condition was state $i$. A straightforward probabilistic argument leads to a set of difference equations obeyed by the fixation probabilities
\begin{equation}
\label{piirecursion}
\pi_i \,=\, \beta_i \pi_{i-1} + (1- \alpha_i-\beta_i) \pi_i + \alpha_i \pi_{i+1} \;.
\end{equation}
If we define the discrete derivative of $\pi_i$ as
\begin{equation}
\label{defderivative}
d_i \,=\, \pi_i-\pi_{i-1} \;,
\end{equation}
then (\ref{piirecursion}) can be rewritten as
\begin{equation}  \label{direcursion}
 d_{i+1} = r_i^{-1} d_i \;,
\end{equation}
 $i=1,2, \dots, N-1$, where
 \begin{equation}  \label{defri}
r_i \,=\, \frac{\alpha_i}{\beta_i} \,=\,\frac{f_i}{g_i}
 \end{equation}
 is just the relative fitness of A individuals with respect to B individuals. As all elements in the pay-off matrix (\ref{payoffmatrix}) are positive, and $w \in [0,1]$, then $r_i>0$ for all $i$.
 
An exact and explicit formula for $\pi_i$ may be obtained by solving equations (\ref{piirecursion}), see e.g. \cite{nowakbook}. The first step for that is solving (\ref{direcursion}), obtaining $d_i$ for $i=2, 3, \dots, N$ in terms of $d_1$. Taking into account the boundary conditions 
 \begin{equation}
 \label{boundcond}
 \pi_0 \,=\, 0 \;\;\;\mathrm{and} \;\;\; \pi_N\,=\,1
 \end{equation}
 and the fact that $\sum_{i=1}^{N} d_i = \pi_N- \pi_{0}=1$, it can be seen that
 \begin{equation}
 \label{pi1formula}
 \pi_1 \,=\, \frac{1}{1+ \sum_{j=1}^{N-1} \prod_{k=1}^j r_k^{-1}}
 \end{equation}
 and, for $i=2, \dots, N-1$,
\begin{equation}
\label{piiformula}
\pi_i \,=\, \frac{1+ \sum_{j=1}^{i-1} \prod_{k=1}^j r_k^{-1}}{1+ \sum_{j=1}^{N-1} \prod_{k=1}^j r_k^{-1}} \;.
\end{equation}

Although exact and explicit, the above formulas are ``unwieldy" \cite{ewens}, unless fitnesses are independent of frequency. The neutral case is the simplest, because $r_k=1$ for all $k$, and we get immediately
\begin{equation}  \label{neutralpii}
\pi_i \,=\, \frac{i}{N}\;.
\end{equation}

A mutant of a new kind in an otherwise uniform population may not reach fixation, even if the mutation is beneficial. In fact, due to the stochastic character of natural selection, as modeled by the Moran process, the mutant and its eventual offspring may be chosen for death in a few time steps. Accordingly, a quantity of interest in Population Genetics is the fixation probability $\rho_A \equiv \pi_1$ of a single mutant of type A in a population of $N-1$ individuals of type B, and also its analogue  $\rho_B \equiv 1-\pi_{N-1}$. For the neutral Moran process, by (\ref{neutralpii}), we have 
\begin{equation}  \label{neutralrho}
\rho_A=\rho_B=1/N\;.
\end{equation}

Classification of the evolutionary scenarios for the Moran process for a population of types A and B was performed by \cite{taylor} using the signs of $r_1-1$, $r_{N-1}-1$, $\rho_A-1/N$ and $\rho_B-1/N$. As stated in the above cited paper, the former two signs characterize the \textit{invasion dynamics}, whereas the latter two characterize \textit{replacement dynamics}.

For instance, $r_1>1$ means that a single A in a population of $N-1$ B individuals is fitter than the Bs. In this case we will say that \textit{natural selection favors the invasion of the B population by a single A mutant}. Using a notation introduced in \cite{taylor} and used again in \cite{AntalScheuring}, $r_1>1$ is denoted as $B\stackrel{\rightarrow \;\;}{\;\;} A$.

Not only we are concerned with comparing the fitness of a single mutant with the rest of the population, but also we will compare the fixation probability of this mutant with the fixation probability  (\ref{neutralrho}) of the mutant in the neutral case. If $\rho_A>1/N$, then we will say that \textit{natural selection favors the replacement of B by A}. The notation for that is $B\stackrel{}{\Rightarrow \;} A$.

Following \cite{taylor} and also \cite{AntalScheuring}, the notation for evolutionary scenarios in the present paper uses single arrows at the top to refer to invasion dynamics, i.e. fitness comparisons, and double arrows at the bottom to refer to replacement dynamics, i.e. fixation probability comparisons with respect to the neutral case. One example to illustrate: the case $r_1>1$, $r_{N-1}<1$, $\rho_A>1/N$, $\rho_B <1/N$, i.e., \textit{selection favors invasion of B by A and invasion of A by B, favors replacement of B by A, but opposes replacement of A by B} is denoted as $B\stackrel{\rightarrow \leftarrow}{\Rightarrow \Rightarrow} A$.

Each combination of arrow orientations in the above notation, or, equivalently, each of the sign possibilities for $r_1-1$, $r_{N-1}-1$, $\rho_A-1/N$ and $\rho_B-1/N$ is defined to be an \textit{evolutionary} scenario for the Moran process with two strategies. Sometimes we will also take into account only the top arrows in the notation and refer to that as an \textit{invasion} scenario. 

Disregarding the exceptional scenarios in which the above quantities may be null, we have 4 arrows with two possible orientations for each, amounting to 16 evolutionary scenarios. We consider the following as the most important result in \cite{taylor}:
\begin{theorem}\label{theoevoscenarios}
	Among the 16 combinatorially possible evolutionary scenarios for the Moran process with two strategies, only 8 may occur:
	\begin{equation}  \label{dominancescenarios}
B\stackrel{\rightarrow \rightarrow}{\Rightarrow \Rightarrow} A \;\;\;\mathrm{and}\;\;\; B\stackrel{\leftarrow \leftarrow}{\Leftarrow \Leftarrow} A\;,
	\end{equation}
	\begin{equation}  \label{mutualscenarios}
B\stackrel{\rightarrow \leftarrow}{\Rightarrow \Rightarrow} A \;, B\stackrel{\rightarrow \leftarrow}{\Leftarrow \Leftarrow} A \;\;\;\mathrm{and}\;\;\; B\stackrel{\rightarrow \leftarrow}{\Rightarrow \Leftarrow} A\;,
	\end{equation}
	and finally 
\begin{equation}   \label{bistabscenarios}
B\stackrel{\leftarrow \rightarrow}{\Rightarrow \Rightarrow} A, B\stackrel{\leftarrow \rightarrow}{\Leftarrow \Leftarrow} A \;\;\;\mathrm{and}\;\;\; B\stackrel{\leftarrow \rightarrow}{\Leftarrow \Rightarrow} A \;.
\end{equation}
\end{theorem}

A new proof of this result, simpler than the one in \cite{taylor},  can be found in \ref{proofevoscenarios}.

\section{Graph shapes}
\label{secgraphshapes}
We start with two simple auxiliary results.
\begin{lemma}  \label{monotoniclemma}
The relative fitness defined by (\ref{defri}) is either a monotonic function of $i$, or it is independent of $i$.
\end{lemma}

We leave to the reader the easy proof of the above result. It suffices to take the derivative of (\ref{defri}) with respect to $i$ and notice that its sign is independent of $i$.

\begin{lemma}  \label{telescopiclemma}
		If $d_i>1/N$ for all $i > j$ and $\pi_j \geq \frac{j}{N}$, then $\pi_i > \frac{i}{N}$ for all $i > j$.
\end{lemma}

\textbf{Proof}
	Notice that if $i>j$, then $\pi_i$ can be written as a telescopic sum
	\[ \pi_i \,=\, \pi_j+ d_{j+1}+ \dots + d_i \;. \]
	By the hypotheses, the first term in the right-hand side is not smaller than $j/N$ and all the $i-j$ remaining terms are larger than $1/N$. It follows that $\pi_i> j/N+\frac{i-j}{N} =i/N$. $\blacksquare$

We consider now the graph shape of the fixation probability in scenario $B\stackrel{\leftarrow \leftarrow}{\Leftarrow \Leftarrow} A$. As the notation suggests, in this case B dominates A. From the point of view of the invasion dynamics, this is a simple consequence of Lemma \ref{monotoniclemma}. In fact, because $r_1<1$ and $r_{N-1}<1$, then no matter whether $r_i$ is increasing, decreasing or constant, we must have $r_i<1$ for all $i$. This means that B is fitter than A no matter the frequency of A in the population. On the other hand, this fact does not automatically mean that the fixation probability of A is smaller than in the neutral case (\ref{neutralpii}) for all frequencies of A. The double arrows in $B\stackrel{}{\Leftarrow \Leftarrow} A$ mean $\rho_A<1/N$ and $\rho_B>1/N$, which imply that for $i=1$ and $i=N-1$ we have indeed $\pi_i<i/N$. That this relation holds for all other values of $i$ is part of the content of the following Proposition.

\begin{figure*}
	\includegraphics[width=  \textwidth]{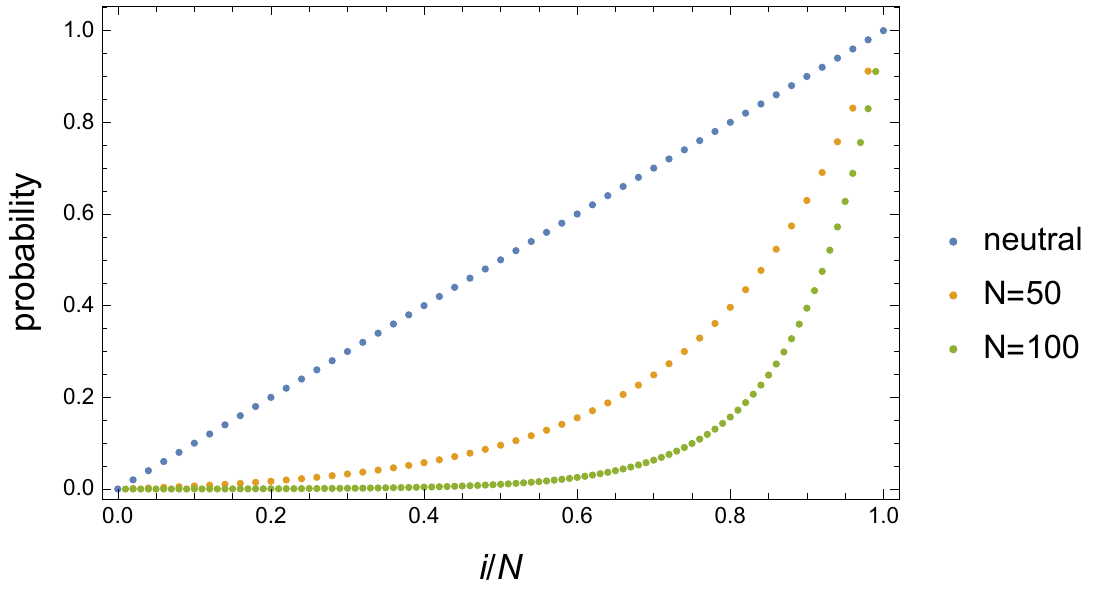}
      % Give a unique label
      \caption{Plot of the fixation probability of A as a function of the initial fraction $i/N$ of A individuals in the population for two values of the population size. The neutral case graph $\pi_i=i/N$ is shown for comparison. The pay-off matrix elements are $a=	1$, $b=1.1$, $c=1.1$ and $d=1.2$  and the selection intensity is $w=1$. For both values $N=50$ and $N=100$ it is easy to check that $r_1<1$ and $r_{N-1}<1$. From the graphs it is seen that conditions $\rho_A<1/N$ and $\rho_B>1/N$ are fulfilled, so that the evolutionary scenario is $B\stackrel{\leftarrow \leftarrow}{\Leftarrow \Leftarrow} A$.}
      \label{figdominanceshape}
\end{figure*}

\begin{proposition}   \label{dominanceshape}
	In scenario $B\stackrel{\leftarrow \leftarrow}{\Leftarrow \Leftarrow} A$ we have $\pi_i<i/N$ for all $i \in \{1, 2, \dots, N-1\}$. Moreover the discrete derivative $d_i=\pi_i-\pi_{i-1}$ is an increasing function of $i$.
\end{proposition}
The proof of Proposition \ref{dominanceshape} is in \ref{proofdominanceshape}.  Figure \ref{figdominanceshape} illustrates some graph shapes with the characteristics proved in the above result.

If, instead of focusing on the fixation probability of A, we consider the fixation probability of B, then we are led to define $\overline{\pi}_i$ the fixation probability of type B when the initial number of B individuals in the population is $i$. Of course, 
\begin{equation}  \label{pibar}
\overline{\pi}_i \,=\, 1-\pi_{N-i} \;.
\end{equation}
Using this duality, the reader may prove a result similar to Proposition \ref{dominanceshape} above for the scenario $B\stackrel{\rightarrow \rightarrow}{\Rightarrow \Rightarrow} A$:
\begin{proposition}   \label{dualdominanceshape}
	In scenario $B\stackrel{\rightarrow \rightarrow}{\Rightarrow \Rightarrow} A$ we have $\pi_i>i/N$ for all $i \in \{1, 2, \dots, N-1\}$. Moreover the discrete derivative $d_i=\pi_i-\pi_{i-1}$ is a decreasing function of $i$.
\end{proposition}

\begin{figure*}
	\includegraphics[width=  \textwidth]{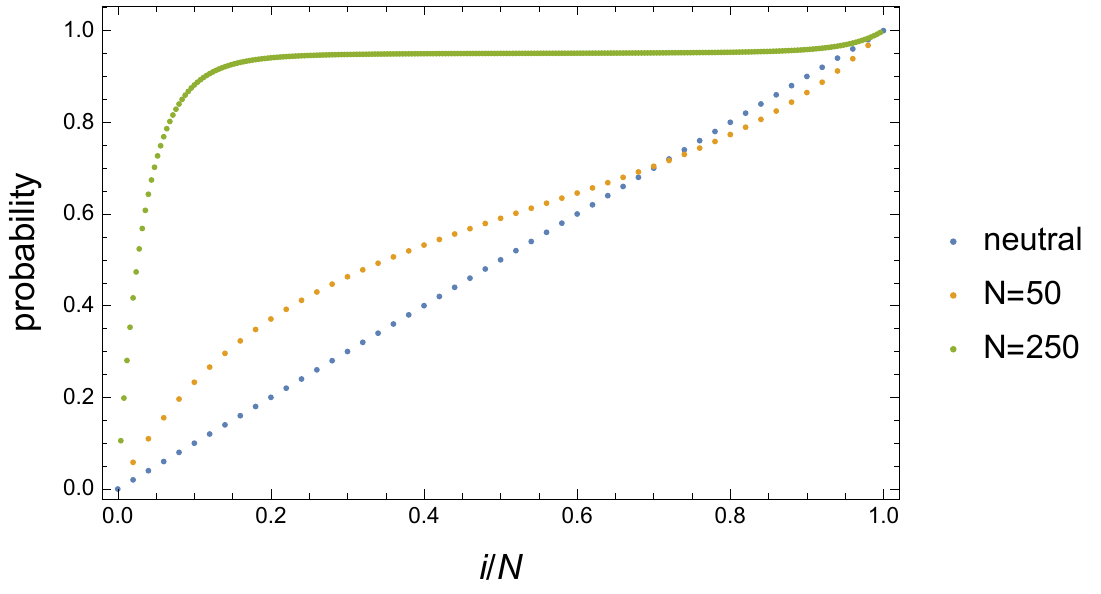}
	\caption{Plot of the fixation probability of A as a function of the initial fraction $i/N$ of A individuals in the population for two values of the population size. The neutral case graph $\pi_i=i/N$ is shown for comparison. The pay-off matrix elements are $a=2$, $b= 2.1$, $c=2.2$ and $d= 1.85$  and the selection intensity is $w=1$.
	For both values $N=50$ and $N=250$ it is easy to check that $r_1>1$ and $r_{N-1}<1$. For $N=50$ we have $\rho_A>1/N$ and $\rho_B>1/N$ and the evolutionary scenario is $B\stackrel{\rightarrow \leftarrow}{\Rightarrow \Leftarrow} A$. But for $N=250$ we have $\rho_A>1/N$ and $\rho_B<1/N$, so that the evolutionary scenario is $B\stackrel{\rightarrow \leftarrow}{\Rightarrow \Rightarrow} A$.}
	\label{figcoexistenceshape}      % Give a unique label
\end{figure*}

Let us now consider the three evolutionary scenarios in Theorem \ref{theoevoscenarios} when the invasion scenario is $B\stackrel{\rightarrow \leftarrow}{} A$, i.e. those in (\ref{mutualscenarios}). Their graph shapes are characterized by the following
\begin{proposition}  \label{coexistenceshape}
	If the invasion scenario is $B\stackrel{\rightarrow \leftarrow}{} A$, there exists a single $i^{\ast} \in \{1, 2, \dots, N-1\}$ such that $d_{i+1}< d_i$ for $i <i^{\ast}$, $d_{i+1}> d_i$ for $i >i^{\ast}$ and $d_{i^{\ast}+1} \geq d_{i^{\ast}}$.	
	
	In scenario $B\stackrel{\rightarrow \leftarrow}{\Rightarrow \Leftarrow} A$ there exists a single $\overline{i} \in \{2, \dots, N-2\}$ such that $\pi_i \geq i/N$ for $i < \overline{i}$ and $\pi_i < i/N$ for $i \geq \overline{i}$.
	
	In scenarios $B\stackrel{\rightarrow \leftarrow}{\Rightarrow \Rightarrow} A$ and $B\stackrel{\rightarrow \leftarrow}{\Leftarrow \Leftarrow} A$ we respectively have $\pi_i>i/N$ and $\pi_i<i/N$ for all $i \in \{1,2, \dots, N-1\}$.
\end{proposition}
Notice that $i^{\ast}$ is a point at which the discrete derivative changes its decreasing behavior to an increasing behavior. We may define $i^{\ast}$ to be an inflection point. Proposition \ref{coexistenceshape} proves existence and uniqueness of an inflection point for the graph in all three scenarios (\ref{mutualscenarios}). On the other hand, $\overline{i}$ is a point where the graph of the fixation probability crosses the neutral line. It exists only in scenario $B\stackrel{\rightarrow \leftarrow}{\Rightarrow \Leftarrow} A$ and is unique. The proof of these statements is in \ref{proofcoexistence}. The graphs in Figure \ref{figcoexistenceshape} illustrate the shape for two among the scenarios $B\stackrel{\rightarrow \leftarrow}{} A$. An example for the remaining scenario may be obtained by exchanging types A and B in the pay-off matrix in the Figure.

\begin{figure*}
	\includegraphics[width=  \textwidth]{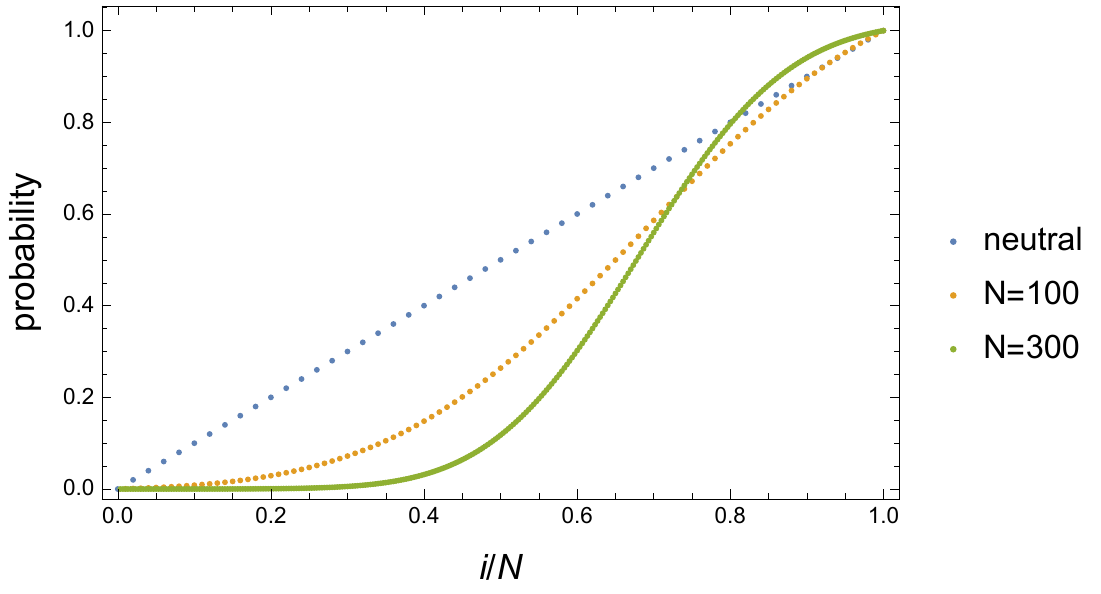}
	\caption{Plot of the fixation probability of A as a function of the initial fraction $i/N$ of A individuals in the population for two values of the population size. The neutral case graph $\pi_i=i/N$ is shown for comparison. The pay-off matrix elements are $a=2.3$, $b= 1.89$, $c=2.2$ and $d= 2.1$ and the selection intensity is $w=1$.
		For both values $N=100$ and $N=300$ it is easy to check that $r_1<1$ and $r_{N-1}>1$. For $N=100$ we have $\rho_A<1/N$ and $\rho_B>1/N$ and the evolutionary scenario is $B\stackrel{\leftarrow \rightarrow}{\Leftarrow \Leftarrow} A$. But for $N=300$ we have $\rho_A<1/N$ and $\rho_B<1/N$, so that the evolutionary scenario is $B\stackrel{\leftarrow \rightarrow}{\Leftarrow \Rightarrow} A$.}
	\label{figbistabshapes}     % Give a unique label
\end{figure*}

Finally, for the invasion scenarios $B\stackrel{\leftarrow \rightarrow}{} A$ in (\ref{bistabscenarios}), the graph shapes are determined by the following result, analogous to Proposition \ref{coexistenceshape}. In all cases there is a single inflection point $i^{\ast}$ and only in $B\stackrel{\leftarrow \rightarrow}{\Leftarrow \Rightarrow} A$ the graph crosses the neutral line, but only once. The proof is omitted, because it follows the same ideas of the proof of the analogous Proposition \ref{coexistenceshape}. The shapes are illustrated in Figure \ref{figbistabshapes}.

\begin{proposition}  \label{coordinationshape}
	If the invasion scenario is $B\stackrel{\leftarrow \rightarrow}{} A$, there exists a single $i^{\ast} \in  \{1, 2, \dots, N-1\}$ such that $d_{i+1}> d_i$ for $i <i^{\ast}$, $d_{i+1}< d_i$ for $i >i^{\ast}$ and $d_{i^{\ast}+1} \leq d_{i^{\ast}}$.	
	
	In scenario $B\stackrel{\leftarrow \rightarrow}{\Leftarrow \Rightarrow} A$ there exists a single $\overline{i} \in \{2, \dots, N-2\}$ such that $\pi_i \leq i/N$ for $i < \overline{i}$ and $\pi_i > i/N$ for $i \geq \overline{i}$.
	
	In scenarios $B\stackrel{\leftarrow \rightarrow}{\Leftarrow \Leftarrow} A$ and $B\stackrel{\leftarrow \rightarrow}{\Rightarrow \Rightarrow} A$ we respectively have $\pi_i<i/N$ and $\pi_i>i/N$ for all $i \in \{1,2, \dots, N-1\}$.
\end{proposition}

\section{Large population asymptotics: statement of the problem}
\label{secasymptotics}
From now on, we will consider the behavior of the fixation probabilities $\pi_i$ when both the population size $N$ and the number of A individuals $i$ tend to infinity, but with the fraction $x=i/N$ held fixed. For this sake, we define for $x \in[0,1]$
\begin{equation}  \label{defPiN}
\Pi_N(x)= \pi_{[Nx]}\;.
\end{equation}

We start with
 the exact formula (\ref{piiformula}). Defining 
\begin{equation}
\label{discpotential}
\ell_j \,=\, - \sum_{k=1}^j \frac{1}{N}\, \log r_k \;,
\end{equation}
we may rewrite (\ref{piiformula}) as
\begin{equation}
\label{piiell}
\pi_i \,=\, \frac{1+ \sum_{j=1}^{i-1} e^{ N \ell_j}}{1+ \sum_{j=1}^{N-1} e^{ N \ell_j}} \;.
\end{equation}
Using (\ref{fi}) and (\ref{gi}) in (\ref{defri}), it is easy to see that
\begin{equation}
\label{convrtoR}
r_{[Nx]} \stackrel{N \rightarrow \infty}{\longrightarrow}  R(y) \;,
\end{equation}
where
\begin{equation}
\label{defR}
R(y) \,\equiv \, \frac{1-w+w[ay+b(1-y)]}{1-w+w[cy+d(1-y)]}\;
\end{equation}
and  $y=[Nx]/N=x+\delta$, with $\delta$ being a remainder term with $|\delta|\le 1/(2N)$. In order to avoid such remainders and some small, but non-negligible, effects they have when multiplied by $N$, when necessary we will demand that the fraction $x$ of A individuals in the population is in $\mathbb{Q}_{N_0}$ for some $N_0 \in \mathbb{N}$ and also that the limit of infinite population $N \rightarrow \infty$ is taken with $N$ being a multiple of $N_0$.

Observe that, due to (\ref{convrtoR}), when $N$ is large enough, the signs of $r_1-1$ and $R(0)-1$ are the same, as well as the signs of $r_{N-1}-1$ and $R(1)-1$. This means that for large enough $N$ the invasion scenario part (i.e. the upper arrows) in the classification of evolutionary scenarios in Theorem \ref{theoevoscenarios} can be rephrased in terms of the signs of $R(0)-1$ and $R(1)-1$.

The reader should also notice that a straightforward analogue of Lemma \ref{monotoniclemma} holds for the above defined function $R$, because its derivative $R'$ is either strictly positive, or strictly negative in $[0,1]$.

Because we will need it afterwards, we make now more precise some duality arguments we had already hinted at in (\ref{pibar}). More precisely, we define $\overline{a}=d$, $\overline{b}=c$, $\overline{c}=b$ and $\overline{d}=a$ and a new pay-off matrix
\begin{equation}  \label{barpayoff}
\overline P \,=\,\left(\begin{array}{cc}
\overline a& \overline b\\
\overline c&\overline d
\end{array}
\right) \,=\, \left(\begin{array}{cc}
d & c\\
b &a
\end{array}
\right) 
\end{equation}
which interchanges A and B individuals. Let $\overline{\pi}_i$ denote the fixation probability for $B$ individuals when the initial number of Bs is $i$. The relationship between $\pi$ and $\overline \pi$ is (\ref{pibar}). Of course we may also calculate $\overline \pi_i$ by
\begin{equation}
\label{pibarell}
\overline \pi_i \,=\, \frac{1+ \sum_{j=1}^{i-1} e^{ N \overline \ell_j}}{1+ \sum_{j=1}^{N-1} e^{ N \overline \ell_j}} \;,
\end{equation}
where $\overline r_j$ and $\overline \ell_j$ are defined as $r_j$ and $\ell_j$ by replacing the pay-offs $a$, $b$, $c$ and $d$ by their barred counterparts. It is easy to see that $\overline r_j= 1/r_{N-j}$. Defining
\begin{equation}
\label{defRbar}
\overline R(y) \,\equiv \, \frac{1-w+w[\overline ay+\overline b(1-y)]}{1-w+w[\overline cy+\overline d(1-y)]}\;,
\end{equation}
it also holds that
\begin{equation}  \label{dualR}
\overline R(\overline y) \,=\, \frac{1}{R(y)}\;,
\end{equation}
where $\overline y=1-y$.

Returning to  (\ref{discpotential}), we see that the sum its right hand side is approximately a Riemann sum, which converges when $N \rightarrow \infty $ to an integral. We may then write the \textit{approximation}
\[\ell_j \approx L(\frac{j}{N}) \;\]
where 
\begin{equation}
\label{defL}
L(x) \,\equiv  \, - \, \int_{0}^{x} \log R(t) \,dt \;.
\end{equation}
The above defined integral, if multiplied by $N$ and another constant, is exactly the \textit{fitness potential} defined in \cite{ChalubSouza2016}. Due to positivity of $a$, $b$, $c$ and $d$, $L$ is a $C^{\infty}$ function on $[0,1]$. The integral may be calculated exactly, but we need not do that. Returning to (\ref{piiell}), it becomes, on using the approximation,
\begin{equation}  \label{approxpii}
\pi_i \, \approx \, \frac{1+ \sum_{j=1}^{i-1} e^{N L(j/N)}}{1+ \sum_{j=1}^{N-1} e^{N L(j/N)}} \;.
\end{equation}

Notice that for large $N$ the sums in the numerator and in the denominator of the preceding formula are dominated by the values of $j$ close to the values in which $L$ is maximum. As we shall see, which values of $j$ maximize $L(j/N)$ depends on which among the four possibilities for the signs of $R(0)-1$ and $R(1)-1$ we are in. 

\cite{AntalScheuring} used (\ref{approxpii}) above as if it were exact and obtained asymptotic results when $N \rightarrow \infty$. Technically, what we will do is a version for sums of the Laplace method, see e.g. \cite{olver}, for asymptotic evaluation of certain integrals, resulting in a rigorous version of the results of the above authors. 

First of all, we define 
\begin{equation}
\label{defsi}
s_i \,=\, \left\{\begin{array}{ll}
1, & i=1 \\
1+ \sum_{j=1}^{i-1} e^{N \ell_j}, & i=2, 3, \dots, N 
\end{array}\;,
 \right.
\end{equation}
so that 
\begin{equation} \label{rhoA}
\rho_A = \pi_1 \,=\, \frac{1}{s_N}
\end{equation}
and
\begin{equation}
\label{PiNsi}
\Pi_N(x) \,=\, \frac{s_{[Nx]}}{s_{N}}\;.
\end{equation}
We will proceed by obtaining asymptotic formulae as $N \rightarrow \infty$ for the numerator and denominator of the above formula. Instead of the approximations which led to (\ref{approxpii}), let us now write exact expressions for the numerator and denominator in (\ref{PiNsi}), separating them in parts we are able to estimate. 

Both numerator and denominator may be rewritten as
\begin{eqnarray}
s_{[Nx]} &=& 1+ \sum_{j=1}^{[Nx]-1} e^{N L(\frac{j}{N})} \,+\, \sum_{j=1}^{[Nx]-1} e^{N L(\frac{j}{N})}\, \left[e^{N(\ell_j-L(\frac{j}{N}))}-1 \right] \label{bigexpsN}\\
&=& 1+ s_{[Nx]}^c+ s_{[Nx]}^m \nonumber\;,
\end{eqnarray}
where
\begin{equation}
\label{defsnc}
s_{[Nx]}^c \,=\, \sum_{j=1}^{[Nx]-1} e^{N L(\frac{j}{N})}\, \left[e^{N(\ell_j-L(\frac{j}{N}))}-1 \right] 
\end{equation}
and
\begin{equation} \label{defsnm}
s_{[Nx]}^m \,=\, \sum_{j=1}^{[Nx]-1} e^{N L(\frac{j}{N})} \;.
\end{equation}

The $s_{[Nx]}^m$ term is the ``main" contribution to $s_{[Nx]}$, the one considered in \cite{AntalScheuring}.
The term $s_{[Nx]}^c$, on the other hand, is due to having replaced $\ell_j$ by its continuous approximation $L(\frac{j}{N})$. It is an instance of what we have called ``continuation error" in Section \ref{secintro}.  

Before we embark in studying the asymptotics for the fixation probabilities in the particular invasion scenarios, we introduce a result which will be useful in dealing with the continuation error terms. At first, one could naively think that they tend to 0 as $N \rightarrow \infty$, because $\ell_j-L(j/N)$ tends to 0 in this limit. But we cannot  conclude that the term between brackets in (\ref{defsnc})  tends to 0. In fact, we can show that $N(\ell_j-L(\frac{j}{N}))$ is bounded and has a nonzero limit when $N \rightarrow \infty$. More precisely:
\begin{proposition}  \label{propconvq}
Let
\begin{equation}  \label{defQ}
Q(x)\,=\, \frac{1}{2}\, \log \frac{R(0)}{R(x)}- w \, q(x) \;,
\end{equation}
where 
\begin{equation}  \label{defq}
q(x) \,=\, \int_0^x \left[\frac{d}{1-w+w(ct+d(1-t))}- \frac{a}{1-w+w(at+b(1-t))} \right] \, dt \;.
\end{equation}
Let also  $x \in [0,1] \cap \mathbb{Q}_{N_0}$ for some $N_0 \in \mathbb{N}$. If $N=m N_0$, $m \in \mathbb{N}$, then $N(\ell_{[Nx]}-L(x))$ converges as $m \rightarrow \infty$ to $Q(x)$.
\end{proposition}	

The proof for the above Proposition can be seen in \ref{subpropconvq}.

We will start the analysis of the particular invasion scenarios by the coordination game scenario $B\leftarrow \rightarrow A$. It will be seen that this is the most difficult scenario, because it is the only one in which the maximum of $L$ occurs at an interior point of $[0,1]$.

\section{The coordination case $B\leftarrow \rightarrow A$}\label{seccoord}
 For large enough $N$, $B\leftarrow \rightarrow A$ is characterized by  $R(0)<1$ and $R(1)>1$. Because, as already remarked, the derivative of $R$ has a fixed sign in $[0,1]$, $R$ must then be an increasing function in this scenario. As a consequence, the function $L$ defined by (\ref{defL}) must have a maximum at the point $x^{\ast}$ at which $R$ passes through the value 1. Of course, $L'(x^{\ast})=0$ and $L''(x^{\ast})= -R'(x^{\ast})<0$.

Let $J^{\ast}= [N x^{\ast}]$ and
\begin{equation}
\label{yN}
\delta_N \,=\, N x^{\ast}-J^{\ast} \;,
\end{equation}
so that $|\delta_N| \leq 1/2$. 

We will split both numerator and denominator in (\ref{PiNsi}) as in (\ref{bigexpsN}). We will start by estimating the main part $s_N^m$ of the denominator, that may be written as in (\ref{defsnm}) with $x=1$. As $J^{\ast}<N$, for large enough $N$ we may separate the sum defining $s_N$ into the term with $j=J^{\ast}$ and the terms with $j<J^{\ast}$ and $j>J^{\ast}$. If we change summation indices in the latter two, we get
\begin{eqnarray}
s_N^m&=&e^{NL(x^{\ast})} \, \left[ e^{N(L(\frac{J^{\ast}}{N})-L(x^{\ast}))} + \rule{0cm}{0.6cm} \right. \nonumber\\
&&\left. \sum_{k=1}^{J^{\ast}-1} e^{N(L(\frac{J^{\ast}-k}{N})-L(x^{\ast}))} + \sum_{k=1}^{N-J^{\ast}-1} e^{N(L(\frac{J^{\ast}+k}{N})-L(x^{\ast}))}\right]\;.\label{bigexpsnm}
\end{eqnarray} 

The first term inside the brackets in (\ref{bigexpsnm}) is irrelevant in the limit $N \rightarrow \infty$ when compared with the other two. In fact, the latter two will be seen to tend to $\infty$, whereas the former tends to 1 in the same limit. This is true because, by using (\ref{yN}), we have
\[L(\frac{J^{\ast}}{N})\,=\, L(x^{\ast}-\frac{\delta_N}{N}) \;.\]
Using a second-degree Taylor polynomial with a remainder in Lagrange form, we may write
\begin{eqnarray}  \label{taylorexp3}
&&L(x^{\ast}-\frac{k+\delta_N}{N})\,=\, L(x^{\ast}) \,-\, \frac{1}{2!}\,R'(x^{\ast}) \frac{(k+\delta_N)^2}{N^2}\\
&-& \frac{1}{3!}\,L'''(x^{\ast}- \alpha_N \frac{k+\delta_N}{N})\, \frac{(k+\delta_N)^3}{N^3}\nonumber\;,
\end{eqnarray}
where $\alpha_N$ is some number in $(0,1)$ and we also used that $L'(x^{\ast})=0$ and $L''(x^{\ast})=-R'(x^{\ast})$. Setting $k=0$, it follows that
\[N(L(\frac{J^{\ast}}{N})- L(x^{\ast})) \,=\, -\frac{R'(x^{\ast})}{2N}  \delta_N^2 \,-\, \frac{L'''(x^{\ast}-\frac{\alpha_N \delta_N}{N})}{6N^2} \, \delta_N^3\;.\]
Our claim that $e^{N(L(\frac{J^{\ast}}{N})-L(x^{\ast}))}$ tends to 1 as $N \rightarrow \infty$ follows then as a consequence of the boundedness of $L'''$ in $[0,1]$ and the fact that $|\delta_N| \leq 1/2$.

Due to $L''(x^{\ast})$ being negative, i.e. the approximate symmetry around a quadratic maximum, then the last two terms inside the square brackets in (\ref{bigexpsnm}) will be shown to be asymptotically equal to each other when $N \rightarrow \infty$. In Theorem \ref{theomainasymptotics}, one of our most important results, we will calculate their asymptotic limits. 

\begin{theorem}
	\label{theomainasymptotics}
	In scenario $B\leftarrow \rightarrow A$,
	\begin{equation}
	\label{asympformsnm}
	s_N^m \,\stackrel{N \rightarrow \infty}{\sim} \, \left(\frac{2\pi N}{R'(x^{\ast})}\right)^{\frac{1}{2}} \, e^{N L(x^{\ast})} \;.
	\end{equation}
\end{theorem}

\textbf{Proof}
Using expression (\ref{bigexpsnm}) for $s_N^m$ we have already proved that the first term in the square brackets tends to 1. The result will follow by proving that 
\[
\sum_{k=1}^{J^{\ast}-1} e^{N(L(\frac{J^{\ast}-k}{N})-L(x^{\ast}))}  \,\stackrel{N \rightarrow \infty}{\sim} \, \frac{1}{2} \, \left(\frac{2\pi N}{R'(x^{\ast})}\right)^{\frac{1}{2}}
\]
and also
\[
\sum_{k=1}^{N-J^{\ast}-1} e^{N(L(\frac{J^{\ast}+k}{N})-L(x^{\ast}))}  \,\stackrel{N \rightarrow \infty}{\sim} \, \frac{1}{2} \, \left(\frac{2\pi N}{R'(x^{\ast})}\right)^{\frac{1}{2}}\;.
\]
The proof for both asymptotic formulae above is the same. We will detail it for the first one. 

As we are looking at the limit $N \rightarrow \infty$ and $J^{\ast}$ is $O(N)$, we may suppose that $N$ is so large that $J^{\ast} > N^{2/3}+1$. We write
\[\sum_{k=1}^{J^{\ast}-1} e^{N(L(\frac{J^{\ast}-k}{N})-L(x^{\ast}))} \,=\, \sum_{k=1}^{[N^{2/3}]} e^{N(L(\frac{J^{\ast}-k}{N})-L(x^{\ast}))} \,+\, \sum_{k=[ N^{2/3}]+1}^{J^{\ast}-1} e^{N(L(\frac{J^{\ast}-k}{N})-L(x^{\ast}))}\;. \]

By Taylor expanding around $x^{\ast}$, we can see that the exponent $N(L(\frac{J^{\ast}-k}{N})-L(x^{\ast}))$ may be approximated for small $k$ by $-\frac{R'(x^{\ast})}{2N} \, k^2$. Then
\begin{eqnarray}
&&\sum_{k=1}^{J^{\ast}-1} e^{N(L(\frac{J^{\ast}-k}{N})-L(x^{\ast}))}\,=\, \sum_{k=1}^{\infty} e^{-\frac{R'(x^{\ast})}{2N} \, k^2} \,-\, \sum_{k=[ N^{2/3}]+1}^{\infty} e^{-\frac{R'(x^{\ast})}{2N} \, k^2}\nonumber\\
&+& \sum_{k=1}^{[ N^{2/3}]} \left[e^{N(L(\frac{J^{\ast}-k}{N})-L(x^{\ast}))}- e^{-\frac{R'(x^{\ast})}{2N} \, k^2}\right]\,+\,
\sum_{k=[ N^{2/3}]+1}^{J^{\ast}-1} e^{N(L(\frac{J^{\ast}-k}{N})-L(x^{\ast}))}\;, \nonumber
\end{eqnarray} 
where, as $R'(x^{\ast})>0$ all the series above are convergent.

We will provide asymptotic estimates for each of the four terms in the right-hand side of the last expression. All of them will involve exchanging the sums for integrals using the Euler-Maclaurin formula, Theorem \ref{EulerMactheo} in \ref{subsecgen}, see also \cite{apostol}. These intermediate results are all proved in \ref{subsecinter}.

The first sum will be dealt with in Proposition \ref{mainpart} and will be shown to be  $O(N^{1/2})$, being the only term giving a non-zero asymptotic contribution. The second and fourth sums are very similar, because in both terms the exponent is quadratic in $k$ and negligible for $k=O(N^{2/3})$ or larger. In Proposition \ref{n23} we prove that both terms tend to 0 when $N \rightarrow \infty$. The third sum is the error term appearing when we replace $L(\frac{J^{\ast}-k}{N})-L(x^{\ast})$ by the first non-zero term in its Taylor expansion. We will show in Proposition \ref{properrorTaylor} that it is bounded, thus negligible with respect to the first sum, when $N \rightarrow \infty$.
$\blacksquare$

The continuation error term $s_N^c$ included in the denominator of (\ref{PiNsi}) may be asymptotically estimated using the same ideas as in Theorem \ref{theomainasymptotics}. Of course the sum in (\ref{defsnc}) with $x=1$ is also dominated by the values of $j$ close to $N x^{\ast}$, but the term within square brackets in that formula will yield an $e^{Q(x^{\ast})}-1$, according to Proposition \ref{propconvq}. The result is
\begin{theorem}
	\label{theocontasymptotics}
	If the invasion scenario is $B\leftarrow \rightarrow A$, then
	\begin{equation}
	\label{asympformsnc}
	s_N^c \,\stackrel{N \rightarrow \infty}{\sim} \, \left(\frac{2\pi N}{R'(x^{\ast})}\right)^{\frac{1}{2}} \,(e^{Q(x^{\ast})}-1)\, e^{N L(x^{\ast})} \;.
	\end{equation}
\end{theorem}

As the ideas used in the proof of this result are a mere repetition of the ones used in proving Theorem \ref{theomainasymptotics}, we will not present them.

Having asymptotically estimated both nontrivial terms in the denominator of (\ref{PiNsi}), we must now consider the numerator in the same expression, which may also be decomposed in its main and continuation error parts as in (\ref{bigexpsN}). We will assume that the point $x$ in which we want to estimate $\Pi_N(x)$ is in $\mathbb{Q}_{N_0}$ for some $N_0$ and that $N$ is a multiple of $N_0$. According to whether $x<x^{\ast}$ or $x>x^{\ast}$ we have different situations regarding the value of the summation index $j$ such that $L(\frac{j}{N})$ is maximum. The final result of combining numerator and denominator of (\ref{PiNsi}) in scenario  $B\leftarrow \rightarrow A$, considering all cases for $x$, is assembled in the following result:
\begin{theorem}
	\label{theocompletecoord}
Define
\begin{equation}  \label{defTheta}
\Theta_N(x) \,=\, \frac{1}{1-R(x)} \, \sqrt{\frac{R(x) R'(x^{\ast})}{2\pi}} \, e^{-w(q(x)-q(x^{\ast}))}\, \frac{e^{N(L(x)-L(x^{\ast}))}}{N^{1/2}} \;.
\end{equation}
If the invasion scenario is $B\leftarrow \rightarrow A$, $x \in \mathbb{Q}_{N_0}$ for some $N_0$ and $N$ is a multiple of $N_0$, then
\begin{equation} \label{Picoord}
\Pi_N(x) \,\sim\, \begin{cases} \Theta_N(x), &\mbox{if } x< x^{\ast} \\ 
\displaystyle 1+ \Theta_N(x), & \mbox{if } x>x^{\ast} \end{cases}\;.
\end{equation}

Moreover,
\begin{equation}  \label{rhoAcoord}
\rho_A \sim \left(\frac{R'(x^{\ast})}{2 \pi R(0)  N}\right)^{\frac{1}{2}} e^{-wq(x^{\ast})} \, e^{-NL(x^{\ast})}
\end{equation}
and
\begin{equation}\label{rhoBcoord}
\rho_B \sim \left(\frac{R'(x^{\ast})R(1)}{2 \pi  N}\right)^{\frac{1}{2}} e^{-w(q(x^{\ast})-q(1))} \, e^{-N(L(x^{\ast})-L(1))}\;.
\end{equation}
\end{theorem}
\textbf{Proof}
The sum of contributions estimated in Theorems \ref{theomainasymptotics} and \ref{theocontasymptotics} gives us
\begin{eqnarray}
s_N &\stackrel{N \rightarrow \infty}{\sim} & \left(\frac{2\pi N}{R'(x^{\ast})}\right)^{\frac{1}{2}} \,e^{Q(x^{\ast})}\, e^{N L(x^{\ast})} \nonumber\\
&=& \left(\frac{2\pi R(0) N}{R'(x^{\ast})}\right)^{\frac{1}{2}} \,e^{-w q(x^{\ast})}\, e^{N L(x^{\ast})}
\;.\label{asympformsn}
\end{eqnarray}

If $x<x^{\ast}$, then $L$ is increasing in $[0,x]$ and the maximum of $L(j/N)$ among the summands in (\ref{defsnc}) and (\ref{defsnm}) occurs at $j=Nx-1$. Moreover, $L'$ is positive at this maximum. We have
\[s_{Nx}^m\,=\, \sum_{j=1}^{Nx-1}e^{NL(\frac{j}{N})}\,=\, e^{NL(x)} \, \sum_{k=1}^{Nx-1}e^{N(L(x-\frac{k}{N})-L(x))}\;.\]
Using the Taylor expansion of $L$ centered at $x$, we see that the leading contribution to $N(L(x-\frac{k}{N})-L(x))$ is $-k L'(x)= k \log R(x)$. This suggests us to write
\begin{eqnarray}
s_{Nx}^m&=& e^{NL(x)} \,\left\{\sum_{k=1}^{\infty}R(x)^k \right. \,-\, \sum_{k=Nx}^{\infty}R(x)^k \nonumber \\
&+& \left. \sum_{k=1}^{Nx-1}R(x)^k\left[e^{N(L(x-\frac{k}{N})-L(x))-k \log R(x)}-1 \right]\rule{0cm}{0.5cm} \right\}\;.\nonumber
\end{eqnarray}
As $x<x^{\ast}$, then $0<R(x)<1$ and the first term inside the curly braces in the expression above is a convergent geometric series with sum $\frac{R(x)}{1-R(x)}$. The second term is also a convergent geometric series and can be easily seen to tend to 0 when $N \rightarrow\infty$. Proposition \ref{propsmNx} in \ref{subsecinter} shows that the third term also tends to 0 when $N \rightarrow\infty$. We have thus proved that for $x<x^{\ast}$,
\begin{equation}
s_{Nx}^m \sim \frac{R(x)}{1-R(x)} \, e^{NL(x)} \;.
\end{equation}
With a similar reasoning, and using again Proposition \ref{propconvq}, we have 
\begin{equation}
s_{Nx}^c \sim \frac{R(x)}{1-R(x)} \, (e^{Q(x)}-1)\, e^{NL(x)} \;.
\end{equation}
Summing the above two results and dividing by (\ref{asympformsn}), we get, for $x< x^{\ast}$, $\Pi_N(x) \sim \Theta_N(x)$.

Let now $x>x^{\ast}$. Of course the maximum of $L(j/N)$ occurs now at $j=[N x^{\ast}]$, not at $j=Nx-1$, and all the above calculations break down. To obtain the asymptotic formula for $\Pi_N(x)$ we resort to duality arguments. 

If $\overline{x}=1-x$ and $\overline{x^{\ast}}=1-x^{\ast}$, then $x>x^{\ast}$ implies $\overline{x}< \overline{x^{\ast}}$ and we have $\overline \Pi_N(\overline x) \sim \overline \Theta_N(\overline x)$, where a formula for $\overline \Theta_N(\overline x)$ can be obtained by substituting in (\ref{defTheta}) $R(x)$ by $\overline R(\overline x)$, $L(x)$ by $\overline L(\overline x)$, and so on.

But $\Pi_N(x)=1-\overline \Pi_N(\overline{x})$. So, for $x>x^{\ast}$ we may write
$\Pi_N(x) \sim 1- \overline \Theta_N(\overline{x})$. If in this formula we write all the barred quantities in terms of their unbarred counterparts, using relations such as (\ref{dualR}), we get the second line in (\ref{Picoord}).

Formula (\ref{rhoAcoord}) is obtained by using (\ref{asympformsn}) and $\rho_A=1/s_N$. If we use duality arguments in (\ref{rhoAcoord}) we get (\ref{rhoBcoord}).
$\blacksquare$

Results (\ref{rhoAcoord}) and (\ref{rhoBcoord}) had already appeared at Table 2 in \cite{AntalScheuring}, but with a wrong  numerical factor due to these authors having neglected the continuation error terms. The more general (\ref{Picoord}) does not appear there.

\begin{figure*}
	\includegraphics[width=  \textwidth]{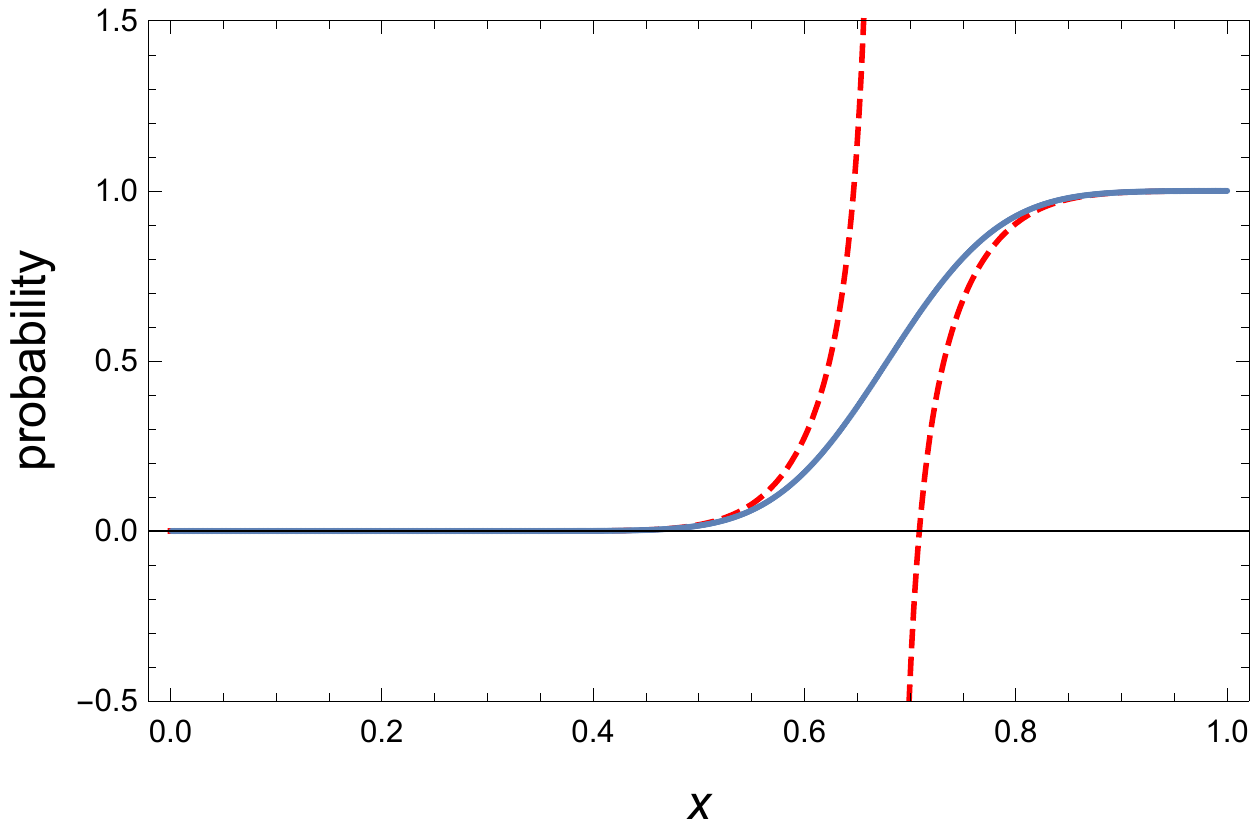}
	\caption{Plots of the fixation probability of A as a function of the initial fraction $x$ of A individuals in the population for $N=1000$ (continuous line) and its asymptotic approximation given by the right-hand side of (\ref{Picoord}) (dashed). The pay-off matrix elements are $a=	2.3$, $b=1.89$, $c=2.2$ and $d=2.1$ and the selection intensity is $w=1$. For this particular matrix, the value of $x^{\ast}$ is approximately $0.68$. The invasion scenario is $B\leftarrow \rightarrow A$.}
	\label{comp1000}      % Give a unique label
\end{figure*}

In Figure \ref{comp1000} we compare for $N=1000$ the numerically computed graph of $\Pi_N$ and its asymptotic approximation given by the right-hand side of (\ref{Picoord}). We see that when $x$ is far from $x^{\ast}$ the two graphs are indistinguishable. Observe that when $x$ is close to $x^{\ast}$ the asymptotic approximation is not only far from good, but not even it yields values in $[0,1]$. 

\begin{figure*}
	\includegraphics[width=  \textwidth]{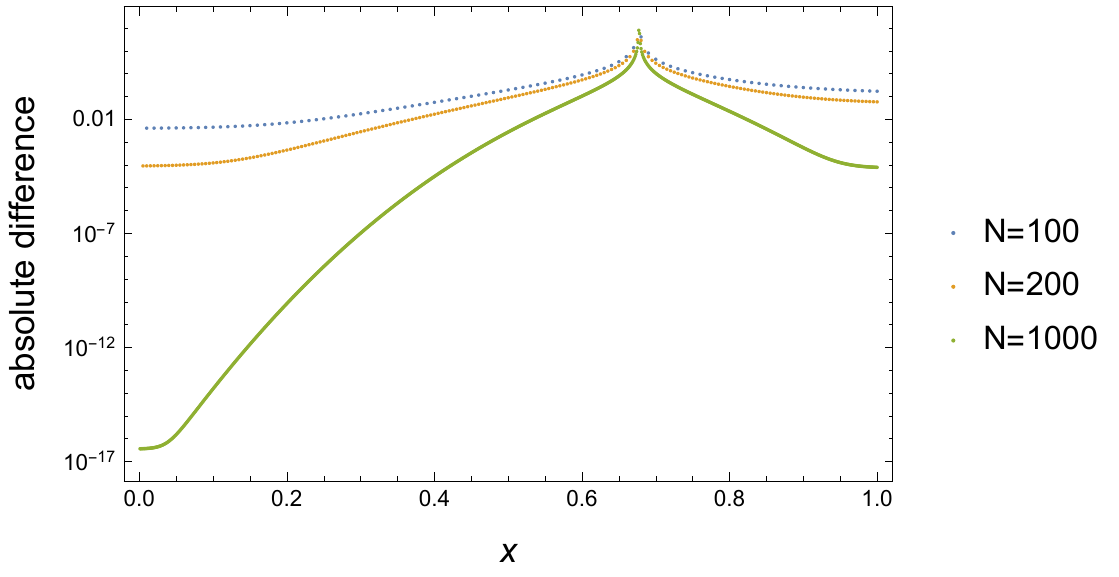}
	\caption{Plots for several values of $N$ of the absolute difference between the fixation probability of A and its asymptotic approximation given by the right-hand side of (\ref{Picoord}). Notice the logarithmic scale on the vertical axis. The pay-off matrix elements are $a=2.3$, $b= 1.89$, $c=2.2$ and $d=2.1$ and the selection intensity is $w=1$. For this particular matrix, the value of $x^{\ast}$ is approximately $0.68$. The invasion scenario is $B\leftarrow \rightarrow A$.}
	\label{complog}      % Give a unique label
\end{figure*}

A better comparison between $\Pi_N$ and its asymptotic approximation is shown in Figure \ref{complog} for several values of $N$. It can be seen that for $x$ far from $x^{\ast}$ the approximation is really good and that the region around $x^{\ast}$ in which it fails decreases as $N$ is increased.

More than a mere approximation scheme, Theorem \ref{theocompletecoord} has the following interesting consequence:
\begin{corollary}
	If the invasion scenario is $B\leftarrow \rightarrow A$, then for large enough $N$ scenarios $B\stackrel{\leftarrow \rightarrow}{\Rightarrow \Rightarrow} A$ and $B\stackrel{\leftarrow \rightarrow}{\Leftarrow \Leftarrow} A$ are ruled out.  
\end{corollary}
\textbf{Proof}
Notice that for $x \in[0,1]$, $x \neq x^{\ast}$, $L(x)<L(x^{\ast})$. Then, for fixed $w$ and $x \neq x^{\ast}$, $\Theta_N(x) \stackrel{N \rightarrow \infty}{\rightarrow}0$. By (\ref{Picoord}), this means that $\Pi_N(x)$ tends to 0 if $x < x^{\ast}$ and to 1 if $x>x^{\ast}$. According to Proposition \ref{coordinationshape}, the only evolutionary scenario compatible with this is  $B\stackrel{\leftarrow \rightarrow}{\Leftarrow \Rightarrow} A$.
$\blacksquare$	

Figure \ref{figbistabshapes} illustrates for a given pay-off matrix the disappearance of scenario $B\stackrel{\leftarrow \rightarrow}{\Leftarrow \Leftarrow} A$ and its substitution for $B\stackrel{\leftarrow \rightarrow}{\Leftarrow \Rightarrow} A$ as the value of $N$ is increased. The importance of the Corollary above is that it reconciles the large $N$ behavior of the Moran process with the deterministic solution of the replicator equation which, in case $B\leftarrow \rightarrow A$, says that B will be fixated if the initial condition is $x<x^{\ast}$ and that A will be fixated if $x>x^{\ast}$.

\section{The remaining invasion scenarios}\label{secremaining}
Consider at first, scenario $B\rightarrow \leftarrow A$,
characterized, for large enough $N$, by  $R(0)>1$ and $R(1)<1$. In this case, $R$ is a decreasing function in $[0,1]$. $L$ has an interior minimum at the point $x^{\ast}$ at which $R$ passes through the value 1, but we already know that what dominates the scenario is the maximum of $L$ in $[0,1]$. This maximum will be at 0 if $L(1)<0$ and at 1 if $L(1)>0$. Of course it may also happen that $L(1)=0$, but we will not consider this non generic case. As the derivative $L'$ does not vanish at the maximum point, the situation will be quite similar to what we found when estimating $s_{Nx}$ for  $x<x^{\ast}$ in the proof of Theorem \ref{theocompletecoord}. The next result synthesizes what we know.
\begin{theorem}  \label{theocoex}
	If the invasion scenario is $B\rightarrow \leftarrow A$, $x \in (0,1) \cap \mathbb{Q}_{N_0}$ for some $N_0$ and $N$ is a multiple of $N_0$, then
	\begin{equation} \label{Picoexist}
	\Pi_N(x) \,\stackrel{N \rightarrow \infty}{\rightarrow}\, \begin{cases} 0, &\mbox{if } L(1)>0 \\ 
	\displaystyle 1, & \mbox{if }  L(1)<0 \end{cases} \;.
	\end{equation}
	
	As a consequence, for large enough $N$ the evolutionary scenario must be $B\stackrel{\rightarrow \leftarrow}{\Leftarrow \Leftarrow} A$ if $L(1)>0$, $B\stackrel{\rightarrow \leftarrow}{\Rightarrow \Rightarrow} A$ if $L(1)<0$. Scenario $B\stackrel{\rightarrow \leftarrow}{\Rightarrow \Leftarrow} A$ is ruled out unless $L(1)=0$.
\end{theorem}
\textbf{Proof}
Consider at first that $L(1)<0$. Then the maximum of $L(x)$ in $[0,1]$ occurs at $x=0$. We have $L'(0)= -\log R(0)<0$ and the first order Taylor expansion around 0 with remainder is $L(j/N)= - \frac{j}{N} \log R(0) + \frac{j^2}{2N^2} L''(\alpha_j \frac{j}{N})$, with $\alpha_j \in(0,1)$. This Taylor expansion suggests writing
\begin{eqnarray}
s^m_{Nx}&=& \sum_{j=1}^{Nx-1}e^{N L(j/N)}\nonumber\\
&=& \sum_{j=1}^{\infty} R(0)^{-j} - \sum_{j=Nx}^{\infty} R(0)^{-j}+ \sum_{j=1}^{Nx-1}R(0)^{-j} \left[e^{N(L(j/N)+\frac{j}{N} \log R(0))}-1\right]\nonumber\\
&=& \frac{1}{R(0)-1} (1-R(0)^{-Nx}) + \sum_{j=1}^{Nx-1}R(0)^{-j} \left[e^{\frac{j^2}{2N} L''(\alpha_j \frac{j}{N})}-1\right]\nonumber\;.
\end{eqnarray}
Using ideas already used in this paper, we see that the last sum tends to 0 when $N \rightarrow \infty$ and we get
\begin{equation}
s^m_{Nx} \stackrel{N \rightarrow \infty}{\rightarrow} \frac{1}{R(0)-1} \;.
\end{equation}
Differently of the other case already studied, we obtain a limit independent of $x$ for $s^m_{Nx}$. 
Using again Proposition \ref{propconvq}, and the fact that $L$ is maximum at 0, we have $s^c_{Nx}\rightarrow \frac{1}{R(0)-1} (e^{Q(0)}-1)=0$, because $Q(0)=0$.

As the above results hold also for $x=1$, the denominator $s_N$ in (\ref{PiNsi}) converges to the same limit as the numerator, proving the second line in (\ref{Picoexist}). 

The first line in (\ref{Picoexist}) is an easy consequence of a reasoning of interchanging A and B as in the proof of Theorem \ref{theocompletecoord}.

Finally, as $\Pi_N(x)$ is close to 1 for all $x>0$, the only compatible scenario if $L(1)<0$, according to Proposition \ref{coexistenceshape}, is $B\stackrel{\rightarrow \leftarrow}{\Rightarrow \Rightarrow} A$. An analogous reasoning proves the statement in case $L(1)>0$.
$\blacksquare$

We observe that, as the continuation error terms in case $B\rightarrow \leftarrow A$ are asymptotically null, then the corresponding results in \cite{AntalScheuring}, shown at their Table 2, are exact.

Figure \ref{figcoexistenceshape} illustrates for a given pay-off matrix the transformation of scenario $B\stackrel{\rightarrow \leftarrow}{\Rightarrow \Leftarrow} A$ for a small value of $N$ to scenario $B\stackrel{\rightarrow \leftarrow}{\Rightarrow \Rightarrow} A$ as the value of $N$ is increased.

We like very much the result of Theorem \ref{theocoex}, because it seems paradoxical. If we expected to find, as in Theorem \ref{theocompletecoord}, some kind of reconciliation between the solution of the deterministic replicator equation for the invasion scenario $B\rightarrow \leftarrow A$ and its Moran counterpart for large $N$, we might be deceived. In fact, the deterministic equation gives stable coexistence of A and B, whereas Theorem \ref{theocoex} gives a criterion, the sign of $L(1)$, to say that either A will fixate certainly, or B will fixate certainly. This result is here to remind us that not always we can expect that stochastic systems converge to their deterministic counterpart as the number of components tends to infinity. We believe that the true reconciliation between stochasticity and determinism in this case is the fact that fixation times in the invasion scenario $B\rightarrow \leftarrow A$ seem to grow very quickly with $N$. Although we present no result on this here, \cite{AntalScheuring} state that this growth is exponential in $N$.

The result for the remaining two invasion scenarios is still easier and we state it without proof:
\begin{theorem}
	If the invasion scenario is $B\rightarrow \rightarrow A$ or $B\leftarrow \leftarrow A$, $x \in (0,1) \cap \mathbb{Q}_{N_0}$ for some $N_0$ and $N$ is a multiple of $N_0$, then
	\begin{equation} \label{Pidominance}
	\Pi_N(x) \,\stackrel{N \rightarrow \infty}{\rightarrow}\, \begin{cases} 0, &\mbox{if scenario is } B\leftarrow \leftarrow A  \\ 
	\displaystyle 1, & \mbox{if scenario is } B\rightarrow \rightarrow A \end{cases} \;.
	\end{equation}
\end{theorem}

\section{Conclusions and outline}\label{secconc}
We have provided in this paper a simple theory which exhibits for each of the eight evolutionary scenarios for the Moran process for two strategies discovered by \cite{taylor} a characteristic shape for the fixation probability $\pi_i$. Although very simple, we had never seen such results before.

In the second part of the paper we showed how these evolutionary scenarios behave when the population size $N$ tends to infinity. This had been done before by \cite{AntalScheuring}, but we wanted to provide a rigorous substitute for their replacement of a sum by an integral, without estimating the error in this exchange. We saw that this error may be null in some cases, but may also amount to a multiplicative factor in another case.

One possible continuation of this study might be studying what happens, if we also take the weak selection limit $w \rightarrow 0$ after limit $N \rightarrow \infty$. Results in this direction stressing that the order in which we perform the two limits is relevant were obtained by \cite{SampleAllen}. From the point of view of this paper, we see that as $w$ is made small, function $L$ becomes almost constant and we are obliged to take larger values of $N$ in order to see the asymptotic behavior shown here.

Another natural continuation is studying the extension for fixation times of the graph shape results and asymptotic behavior when $N \rightarrow \infty$. Some results on that were already obtained by \cite{AntalScheuring}.

\appendix
\section{Proofs}
\label{appproofs}
\subsection{Proof of Theorem \ref{theoevoscenarios}}
\label{proofevoscenarios}

We start with the second scenario in (\ref{dominancescenarios}). We will show that if the invasion dynamics is $B\stackrel{\leftarrow \leftarrow}{} A$, then necessarily $ \rho_A<\frac{1}{N} $ and $ \rho_B>\frac{1}{N} $.  The invasion arrows mean that both $r_1$ and $r_{N-1}$ are smaller than 1. By Lemma \ref{monotoniclemma}, regardless of the $r_i$ being increasing, decreasing or constant, $r_i<1$ for all $i$. This implies $\sum_{j=1}^{N-1}\prod_{i=1}^{j}r_i^{-1}>N-1 $. As $\rho_A=\pi_1$, then (\ref{pi1formula}) implies $\rho_A <1/N$. 

A formula for $\rho_B$ similar to (\ref{pi1formula}) may be obtained: 
\begin{equation*}
\rho_B=\frac{1}{1+\sum_{j=1}^{N-1}\prod_{i=1}^{j}r_{N-i}}\;.
\end{equation*}  
A reasoning similar to the above shows that due to $r_i$ being smaller than 1 for all $i$ implies $\rho_B>1/N$. This proves that the only possible scenario if the invasion dynamics is $B\stackrel{\leftarrow \leftarrow}{} A$ is $B\stackrel{\leftarrow \leftarrow}{\Leftarrow \Leftarrow} A$. An analogous proof holds for the first scenario in (\ref{dominancescenarios}).

Let us now work with the scenarios in (\ref{mutualscenarios}), in which the upper arrows mean $ r_1> 1 > r_{N-1}$. In this case, by Lemma \ref{monotoniclemma}, the $r_i$ are decreasing. Let
\begin{equation*} \ell\equiv \max\{j;r_j\geq1\} \ \ \text{and} \ \ \ell^{\prime}\equiv \max\{j;r_{N-j}\leq1\} \;.
\end{equation*} 
It is clear that $1 \leq \ell \leq N-2$ and the same for $\ell'$. 
We define recursively $ H_i $ as 
\begin{equation*}
H_1=\frac{1}{r_1} \ \text{and} \  H_i=\frac{1}{r_i}H_{i-1}, \ \text{for} \  i=2, 3,\ldots, N-1\;.
\end{equation*}
Analogously, we define $ H^{\prime}_i $ as
\begin{equation*}
H^{\prime}_1=r_{N-1} \ \text{and} \  H^{\prime}_i=r_{N-i}H^{\prime}_i ,\ \text{for} \ i=2,3,\ldots, N-1\;.
\end{equation*}
Notice that sequences $ H_1, H_2,\ldots,H_{\ell} $ and $ H^{\prime}_1, H^{\prime}_2,\ldots,H^{\prime}_{\ell^{\prime}} $ are decreasing, whereas sequences $H_{\ell}, H_{\ell+1},\ldots,H_{N-1} $ and $ H^{\prime}_{\ell^{\prime}}, H^{\prime}_{\ell^{\prime}+1},\ldots,H^{\prime}_{N-1} $ are increasing. See also that $ H_1 < 1 $, $ H^{\prime}_1 < 1 $, but we do not know whether $ H_{N-1} $ and $ H^{\prime}_{N-1} $ are larger, smaller or equal to 1. If $H_i>1$ for some $i$, then we must have $H_{N-1}>1$. If, on the other hand, $H_{N-1}\leq 1$, then $H_i\leq 1$ for all $i$. The same conclusions are valid for the $H'_i$. Moreover, the following relation holds
\begin{equation}\label{hrelation}
H_{N-1}=\frac{1}{H^{\prime}_{N-1}}\;.
\end{equation}
In terms of the new notations, we have  
\begin{equation}\label{hrhoformula}
\rho_A=\frac{1}{1+\sum_{i=1}^{N-1}H_i} \ \ \text{and} \ \ \rho_B=\frac{1}{1+\sum_{i=1}^{N-1}H^{\prime}_i}\;.
\end{equation}

Suppose now that $ \rho_A<1/N $. Then $1/\rho_A-1=\sum_{i=1}^{N-1}H_i>N-1 $ and we must have $H_i>1$ for some $i$. As we have already seen, this implies that $ H_{N-1}>1 $. By relation (\ref{hrelation}), we have $ H^{\prime}_{N-1}<1 $ and thus $ H^{\prime}_i<1 $ for all $ i$. This implies $ \sum_{i=1}^{N-1}H^{\prime}_i<N-1 $ and, by the $ \rho_B $ formula in (\ref{hrhoformula}), we get $ \rho_B>1/N $. 

We can apply the same reasoning above to show that if $ r_1> 1> r_{N-1} $ and $ \rho_B<1/N $, then $ \rho_A>1/N $.

Both reasonings put together show that it is not possible to have both $ \rho_A<1/N $ and $ \rho_B<1/N $ if  $ r_1> 1> r_{N-1} $. In symbols, this means that scenario $B\stackrel{\rightarrow \leftarrow}{\Leftarrow \Rightarrow} A$ is forbidden, proving our assertion about scenarios in (\ref{mutualscenarios}).

The assertion about scenarios (\ref{bistabscenarios}) is proved by an analogous argument. More concretely, we prove that if $ r_1 < 1 < r_{N-1} $, then $ \rho_A>1/N $ and $ \rho_B>1/N $ cannot be both true. In symbols, scenario $B\stackrel{\leftarrow \rightarrow}{\Rightarrow \Leftarrow} A$ is forbidden.$\blacksquare$

Notice that in the above proof we showed that some evolutionary scenarios are forbidden. That all evolutionary scenarios which are not forbidden are actually permitted is illustrated by examples. Examples in Figure \ref{figcoexistenceshape} are a good hint that scenarios $B\stackrel{\rightarrow \leftarrow}{\Rightarrow \Leftarrow} A$ and $B\stackrel{\rightarrow \leftarrow}{\Rightarrow \Rightarrow} A$ are permitted. If the roles of A and B are exchanged, i.e. if the pay-off matrix for the examples in Figure \ref{figcoexistenceshape} is exchanged by $P=\left(
\begin{array}{cc}
1.85 & 2.2 \\
2.1 & 2. \\
\end{array}
\right)$, then we obtain an example of scenario $B\stackrel{\rightarrow \leftarrow}{\Leftarrow \Leftarrow} A$ for $N=250$. Similarly, the examples in Figure \ref{figbistabshapes} are hints that scenarios $B\stackrel{\leftarrow \rightarrow}{\Leftarrow \Leftarrow} A$ and  $B\stackrel{\leftarrow \rightarrow}{\Leftarrow \Rightarrow} A$ do exist. If we exchange A and B, we obtain also an example for the $B\stackrel{\leftarrow \rightarrow}{\Rightarrow \Rightarrow} A$ case.

\subsection{Proofs of the graph shape results}
Before we start, let us introduce some notation for the proofs.

First of all, let $I_{N-1} = \{1,2, \dots, N-1\}$. The set of points $i$ in $I_{N-1}$ such that the fixation probability $\pi_i$ is smaller than the neutral value $i/N$ will be denoted $D$. More concretely,
\[D = \{i \in I_{N-1} \;; \pi_i<i/N\} \;.\]
$\overline{D}$ will then denote a similar set, but replacing the strict inequality by $\leq$, i.e.
\[\overline{D} = \{i \in I_{N-1} \;; \pi_i\leq i/N\} \;.\]
Analogously, we define 
\[U = \{i \in I_{N-1} \;; \pi_i>i/N\} \;\]
and
\[\overline{U} = \{i \in I_{N-1} \;; \pi_i\geq i/N\} \;.\]
\subsubsection{Proof of Proposition \ref{dominanceshape}}
\label{proofdominanceshape}
The scenario hypotheses are $r_1<1$, $r_{N-1}<1$, $\rho_A<1/N$ and $\rho_B>1/N$. The former two imply by Lemma \ref{monotoniclemma} that $r_i<1$ for all $i \in I_{N-1}$. Using (\ref{direcursion}), we prove that the discrete derivative is an increasing function.

The latter two imply that $\pi_1<1/N$ and $\pi_{N-1}<\frac{N-1}{N}$. Using the boundary conditions in (\ref{boundcond}) we also have $d_1<1/N$ and $d_N>1/N$.

What remains to be proved is that $D=I_{N-1}$ in this scenario. We know that $1 \in D$ and $N-1 \in D$. Suppose that $I_{N-1} \setminus D$ is not empty, and let $j$ be its minimum. Then $2 \leq j \leq N-2$, $\pi_{j-1}<\frac{j-1}{N}$ and $\pi_j \geq \frac{j}{N}$. It follows that $d_j>1/N$ and, because the derivative is increasing, $d_i>1/N$ for $i > j$. We may then use Lemma \ref{telescopiclemma} and find that $\pi_{N-1}>\frac{N-1}{N}$, which is an absurd because, as already noticed, we must have $\pi_{N-1}<\frac{N-1}{N}$. Then $I_{N-1} \setminus D$ is empty.
$\blacksquare$

\subsubsection{Proof of Proposition \ref{coexistenceshape}} \label{proofcoexistence}
The arrows in $B\stackrel{\rightarrow \leftarrow}{} A$ mean that we have $r_1>1$ and $r_{N-1}<1$. It is then necessary, according to Lemma \ref{monotoniclemma}, that $r_i$ decreases with $i$. By continuity, equation $r_i=1$ has then a single solution $x \in (1,N-1)$. If we define  $i^{\ast}$ to be the smallest integer larger than or equal to $x$, then using (\ref{direcursion}) we prove the claims related to $i^{\ast}$ in the statement of the proposition.

Suppose now that the scenario is $B\stackrel{\rightarrow \leftarrow}{\Rightarrow \Leftarrow} A$. The double arrows mean $\rho_A>1/N$ and $\rho_B>1/N$, which imply $\pi_1>1/N$ and  $\pi_{N-1}<\frac{N-1}{N}$. So $1 \in U$ and $N-1 \in D$. As $D$ is non-empty, let $\overline{i}$ be its minimum. As $\overline{i} \in D$ and $\overline{i}-1 \in \overline{U}$, then $d_{\overline{i}}<1/N$. 

We claim that $D \supset \{\overline{i}, \overline{i}+1, \dots, N-1 \}$. If this is not true, then $\{\overline{i}, \overline{i}+1, \dots, N-1 \} \setminus D$ is not empty. Let then $j$ be the smallest element in this set. We have $j> \overline{i}$, $\pi_{j-1} < \frac{j-1}{N}$ and $\pi_j \geq \frac{j}{N}$. This implies $d_j>1/N$. Because the discrete derivative increased when going from $\overline{i}$ to $j$, then $j>i^{\ast}+1$. It follows that $d_k> 1/N$ for all $k \geq j$ and, by Lemma \ref{telescopiclemma}, $\pi_{N-1} > \frac{N-1}{N}$, which contradicts hypothesis $\rho_B>1/N$. Then our claim is proved. As $D$ cannot contain elements smaller than its minimum $\overline{i}$, nor larger than $N-1$, then $D = \{\overline{i}, \overline{i}+1, \dots, N-1 \}$, proving what we needed about the scenario $B\stackrel{\rightarrow \leftarrow}{\Rightarrow \Leftarrow} A$.

Consider now $B\stackrel{\rightarrow \leftarrow}{\Rightarrow \Rightarrow} A$. The double arrows here imply $\pi_1>1/N$ and $\pi_{N-1}>\frac{N-1}{N}$. The latter, together with (\ref{boundcond}), implies $d_N<1/N$. What we want to prove is that $\overline{D}=\emptyset$.

Suppose that $\overline{D}$ is not empty. Let then $j_1$ and $j_2$ be its minimum and maximum elements. Because 1 and $N-1$ are in $U$, then $1<j_1 \leq j_2<N-1$. As $j_1-1 \in U$ and $j_1 \in \overline{D}$, then $d_{j_1}<1/N$. Similarly, $d_{j_2+1}>1/N$. As the discrete derivative increased between $j_1$ and $j_2+1$, then $j_2> i^{\ast}$. It follows that $d_j>1/N$ if $j \geq j_2+1$. Together with Lemma \ref{telescopiclemma}, this leads to a contradiction, because we already knew that $d_N<1/N$.

Finally, the proof for scenario $B\stackrel{\rightarrow \leftarrow}{\Leftarrow \Leftarrow} A$ may be obtained from the one of $B\stackrel{\rightarrow \leftarrow}{\Rightarrow \Rightarrow} A$ by using (\ref{pibar}).
$\blacksquare$

\subsection{Proofs of some auxiliary results}\label{secintermediate}

\subsubsection{General results}\label{subsecgen}
For completeness sake, we state here the Euler-Maclaurin formula. For a proof and the definitions of the Bernoulli numbers and periodic functions, the reader is directed to \cite{apostol}.

\begin{theorem}[Euler-Maclaurin formula]\label{EulerMactheo}
For any function $f$ with a continuous derivative of order $2m + 1$ on the interval $[0, n]$, $m \geq 0$ and $n \in \mathbb{N}$, we have
\begin{eqnarray}
\sum_{k=0}^n f(k) &=& \int_0^n f(x)\,dx \,+\, \frac{1}{(2m+1)!}\, \int_0^n P_{2m+1}(x) f^{(2m+1)}(x)\, dx \nonumber\\
&+& \sum_{r=1}^m \frac{B_{2r}}{(2r)!} \, \left[f^{(2r-1)}(n)- f^{(2r-1)}(0)\right] \,+\,\frac{1}{2}(f(n)+f(0))\;, \label{eulerformula}
\end{eqnarray}
where the $P_k$ are the Bernoulli periodic functions and $B_k=P_k(1)$ the Bernoulli numbers.
\end{theorem}

The next general result is a generalized form of the Riemann-Lebesgue lemma appearing in the theory of Fourier series and transform. We state it here for completeness. Its proof will not be presented, because it is basically the proof of the standard form of the same result.

\begin{lemma}[Generalized Riemann-Lebesgue] \label{riemannlebesgue}
Let $p$ be a $C^1$ 1-periodic function with $\int_0^1 p(x)dx=0$ and $f$ be integrable in $[0, \infty)$.
Then
\[\lim_{t \rightarrow \infty} \int_0^{\infty} p(t x) \,f(x) dx \,=\,0 \;.\]
\end{lemma}

\subsubsection{Proof of Proposition \ref{propconvq}}
\label{subpropconvq}
For $x \in [0,1]$ we may write
\[N (\ell_{[Nx]}-L(x)) \,=\, N(\ell_{[Nx]}-L(\frac{[Nx]}{N}))+ N( L(\frac{[Nx]}{N})-L(x))\;.\]
Notice that $|\frac{[Nx]}{N}-x| \leq\frac{1}{2N}$. If we use a Taylor expansion around $x$, it is easy to see that if $x \notin \mathbb{Q}_N$, then the second term in the above sum may be estimated as
\[N( L(\frac{[Nx]}{N})-L(x)) \sim N (\frac{[Nx]}{N}-x) L'(x)\;,\]
which is bounded, but in general does not tend to 0 as $N \rightarrow \infty$. As already remarked, in order to simplify things, we may add the hypothesis that $x\in \mathbb{Q}_{N_0}$ for some $N_0$ and that $N$ is a multiple of $N_0$. With this further assumption, the term we are referring to is null.

Continuing, we will then assume that $Nx$ is integer and we write $\ell_{Nx}$ instead of $\ell_{[Nx]}$. We may split $N (\ell_{Nx}-L(x))$ as
\[N (\ell_{Nx}-L(x)) \,=\, N\left[-\sum_{k=1}^{Nx}\frac{1}{N}\, (\log r_k-\log R(\frac{k}{N})) \,+\, \int_0^x \log R(t)dt \,-\, \frac{1}{N} \, \sum_{k=1}^{Nx} \log  R(\frac{k}{N}) \right] \;,\]
where we used (\ref{defR}) and (\ref{defL}).

For the sum of the second and third terms above, we have
\begin{eqnarray}
N \left[\rule{0cm}{0.5cm} \int_0^x \log R(t)dt \right.&-& \left. \frac{1}{N} \,\sum_{k=1}^{Nx} \log  R(\frac{k}{N}) \right]  \,=\, \int_0^{Nx} \log R(\frac{s}{N}) ds \,-\, \sum_{k=1}^{Nx} \log  R(\frac{k}{N})\nonumber \\&=&\frac{1}{2} (\log R(0) - \log R(x)) \,-\, \int_0^{Nx} P_1(s) \frac{d}{ds} \log R(\frac{s}{N}) \,ds \;, \nonumber 
\end{eqnarray}
where we used  in the last passage the Euler-Maclaurin formula (\ref{eulerformula}) with $m=0$. By making the substitution $u=\frac{s}{N}$ in the last integral and using Lemma \ref{riemannlebesgue}, we prove that it tends to $0$. So, if $Nx$ is an integer,
\[ N \left[ \int_0^x \log R(t)dt \,-\, \frac{1}{N} \,\sum_{k=1}^{Nx} \log  R(\frac{k}{N}) \right] \stackrel{N \rightarrow \infty}{\rightarrow} \frac{1}{2} (\log R(0) - \log R(x))\;.\]

The remaining term $-\sum_{k=1}^{Nx} (\log r_k-\log R(\frac{k}{N}))$ in the above expression for $N (\ell_{Nx}-L(x))$ can be rewritten by using the definitions (\ref{defri}) for $r_k$ and (\ref{defR}) for $R(\frac{k}{N})$. The difference $r_k-R(\frac{k}{N})$ is 
\[r_k-R(\frac{k}{N}) \,=\, \frac{1}{N}\, \frac{d[1-w+w(a\frac{k}{N}+b(1-\frac{k}{N}))]-a[1-w+w(c\frac{k}{N}+d(1-\frac{k}{N}))]}{[1-w+w(c\frac{k}{N}+d(1-\frac{k}{N}))]^2} \,+\, O(\frac{1}{N^2})\;.\]
We may then use the Taylor expansion of the logarithm to find
\begin{eqnarray}
\log r_k &-& \log R(\frac{k}{N}) \,=\, \nonumber \\
&=&\frac{w}{N} \, \left(\frac{d}{1-w+w(c\frac{k}{N}+d(1-\frac{k}{N}))}- \frac{a}{1-w+w(a\frac{k}{N}+b(1-\frac{k}{N}))}\right) \,+\, O(\frac{1}{N^2})\;.\nonumber
\end{eqnarray}
Summing the above expression for $k$ running from 1 to $Nx$, the  first terms become a Riemann sum that converges when $N \rightarrow \infty$ to the integral $w q(x)$, whereas the sum of the $O(1/N^2)$ terms of course tends to 0.$\blacksquare$

\subsubsection{Proofs of some results appearing in Theorems \ref{theomainasymptotics} and \ref{theocompletecoord}}  \label{subsecinter}

We start with the result leading to the only asymptotically non-vanishing contributions for $\sum_{j=1}^{N-1} e^{N L(\frac{j}{N})}$, as in the proof of Theorem \ref{theomainasymptotics}. In fact, these contributions are obtained by just putting $a= R'(x^{\ast})$ in the result below:
\begin{proposition}
\label{mainpart}
Let $a>0$. Then, for any non-negative integer $m$,
\begin{equation} \label{quadsum}
\sum_{k=0}^{\infty} e^{-\frac{a}{N}k^2} \,=\, \frac{1}{2}\, \left(\frac{\pi N}{a}\right)^\frac{1}{2} \,+\, \frac{1}{2} \,+\, o(N^{-m}) \;.
\end{equation}
\end{proposition}
\textbf{Proof}
Take $f(x)=e^{-\frac{a}{N}x^2}$ in (\ref{eulerformula}). We may also take $n=\infty$ in the same formula, because the improper integrals converge and so do the limits at infinity of $f$ and its derivatives, all equal to 0. The odd-ordered derivatives of $f$ at 0 all vanish, too. The Euler-Maclaurin formula gives us then, in this case,
\begin{eqnarray}
\sum_{k=0}^{\infty} e^{-\frac{a}{N}k^2} &=& \int_0^{\infty} e^{-\frac{a}{N}x^2} dx \,+\, \frac{1}{2} f(0)\,+\, \int_0^{\infty} P_{2m+1}(x)\, f^{(2m+1)}(x)\, dx \nonumber\\
&=&  \frac{1}{2}\, \left(\frac{\pi N}{a}\right)^\frac{1}{2} \,+\, \frac{1}{2} \,+\, \int_0^{\infty} P_{2m+1}(x)\, f^{(2m+1)}(x)\, dx \label{eulermacquad}\;.
\end{eqnarray}
To finish the proof, we use the fact that the Hermite polynomials $H_n(x)$ are related to the derivatives of $e^{-x^2}$ by
\[\frac{d^n}{dx^n} \,e^{-x^2} \,=\, (-1)^n \, e^{-x^2} \, H_n(x)\;.\]
The derivatives of $f$ become
\[f^{(n)}(x) \,=\, \left(\frac{a}{N}\right)^{n/2} (-1)^n \, e^{-\frac{a}{N}x^2} \, H_n(\sqrt{\frac{a}{N}}x)\;.\]
Substituting this expression in the remaining integral in (\ref{eulermacquad}), performing the change of variables $u=\sqrt{\frac{a}{N}}x$ and using Lemma \ref{riemannlebesgue}, the result is proved.
$\blacksquare$

Proposition \ref{mainpart} tells us that exchanging the sum in the left-hand side of (\ref{quadsum}) by the corresponding integral produces an error very close to $1/2$. The difference between this error and $1/2$ is so small that for any $m>0$, this difference multiplied by $N^m$ still tends to 0. For the sake of proving Theorem \ref{theomainasymptotics}, $m=1$ would be enough, but our result is so remarkable that we could not help stating it in its full generality.

In the next lemma we prove some asymptotic estimates -- obtained also by the Euler-Maclaurin formula -- necessary for proving Theorem \ref{theomainasymptotics}.
\begin{lemma}  \label{lemmaasympeuler}
\begin{itemize}
\item[(i)] Let $a>0$ and $A \in \mathbb{N}$ be such that $\frac{A^2}{N} \stackrel{N \rightarrow \infty}{\rightarrow} \infty$. Then
	\begin{equation}  \label{asympsumkainf}
		\sum_{k=A}^{\infty} e^{-\frac{a}{N}k^2} \stackrel{N \rightarrow \infty}{\sim}  \left(\frac{1}{2}\, \frac{N}{a} \, A^{-1} \,+\,1\right) \, e^{-\frac{a}{N}A^2}\;. 
	\end{equation}
\item[(ii)] If $p$ is an odd positive integer, then
\begin{equation}  \label{asympsumkp}
\sum_{k=1}^{\infty} k^p \, e^{-\frac{a}{N}k^2} \stackrel{N \rightarrow \infty}{\sim} \frac{1}{2} \, (\frac{p-1}{2})! \, \left(\frac{N}{a}\right)^{\frac{p+1}{2}}\;.
\end{equation}
\end{itemize}
\end{lemma}
\textbf{Proof}
In order to prove (\ref{asympsumkainf}) we use (\ref{eulerformula}), again with $f(x)=e^{-\frac{a}{N} x^2}$ and $n=\infty$, but now we take $m=0$. It gives us
\begin{equation}  \label{sumkainf}
\sum_{k=A}^{\infty} e^{-\frac{a}{N} k^2} \,=\, \int_A^{\infty} e^{-\frac{a}{N} x^2}\,dx \,+\, \int_A^{\infty} P_1(x) \, \frac{d}{dx} e^{-\frac{a}{N} x^2} \,dx \,+\, \frac{1}{2} e^{-\frac{a}{N} A^2}\;.
\end{equation}

For the first integral in the above expression, by a simple change of variable we have
\[\int_A^{\infty} e^{-\frac{a}{N} x^2}\,dx \,=\, \frac{1}{2}\left(\frac{N}{a}\right)^{1/2} \, \int_{\frac{a A^2}{N}}^{\infty} y^{-1/2} \, e^{-y} \, dy \,=\, \frac{1}{2}\left(\frac{N}{a}\right)^{1/2} \,\Gamma(\frac{1}{2}, \frac{a A^2}{N})\;, \]
where $\Gamma(s,x)\equiv \int_x^{\infty} y^{s-1} \, e^{-y}dy$ is the complementary (or upper) incomplete Gamma function. It is known that \cite{olver} $\Gamma(s,x) \stackrel{x \rightarrow \infty}{\sim} x^{s-1} e^{-x}$. Then
\[\int_A^{\infty} e^{-\frac{a}{N} x^2}\,dx \, \stackrel{N \rightarrow \infty}{\sim}\,  \frac{1}{2} \, \frac{N}{a} \, A^{-1} \, e^{-\frac{a}{N} A^2}\;.\]

For the second integral in the right-hand side of (\ref{sumkainf}), we remind that $|P_1(x)| \leq 1/2$ for all $x \in \mathbb{R}$. Thus
\[\left|\int_A^{\infty} P_1(x) \, \frac{d}{dx} e^{-\frac{a}{N} x^2} \,dx\right| \,\leq \, \frac{1}{2} \, \int_A^{\infty} \frac{2a}{N} \,x \, e^{-\frac{a}{N} x^2}\, dx \,=\, \frac{1}{2} \, e^{-\frac{a}{N}A^2}\;.\]

Putting together the expressions for both integrals, we obtain (\ref{asympsumkainf}).

For proving (\ref{asympsumkp}) we use again (\ref{eulerformula}) with $m=0$, now with $f(x)=x^p e^{-\frac{a}{N} x^2}$. We get
\begin{eqnarray}
\sum_{k=1}^{\infty} k^p \, e^{-\frac{a}{N}k^2} &=& \sum_{k=0}^{\infty} k^p \, e^{-\frac{a}{N}k^2} \,=\, \int_0^{\infty} x^p \, e^{-\frac{a}{N} x^2} \, dx \,+\, \int_0^{\infty} P_1(x)\, f'(x) \, dx \,+\, \frac{1}{2}f(0)\nonumber\\
&=& \frac{1}{2} \, \left(\frac{N}{a}\right)^{\frac{p+1}{2}}\, \Gamma(\frac{p+1}{2})\,+\, \int_0^{\infty} P_1(x)\, f'(x) \, dx\;. \nonumber
\end{eqnarray}
Explicitly calculating $f'(x)$ and using again $|P_1(x)|\leq 1/2$, the integral in the right-hand side may be easily bounded by a sum of two terms, both of them being $O(N^{p/2})$, thus negligible with respect to the first term in the right-hand side of the above formula. Using that $\Gamma(\frac{p+1}{2})= (\frac{p-1}{2})!$ for odd $p$ takes us to the result.
$\blacksquare$

The next result proves that two of the terms appearing in the proof of Theorem \ref{theomainasymptotics} vanish asymptotically.
\begin{proposition}  \label{n23}
Both terms
	\[\sum_{k=[ N^{2/3}]+1}^{\infty} e^{-\frac{R'(x^{\ast})}{2N} \, k^2}\]
	and 
	\[
	\sum_{k=[ N^{2/3}]+1}^{J^{\ast}-1} e^{N(L(\frac{J^{\ast}-k}{N})-L(x^{\ast}))}
	\]
	appearing in the proof of Theorem \ref{theomainasymptotics} tend to 0 when $N \rightarrow \infty$.
\end{proposition}

\textbf{Proof}

Using $a=\frac{R'(x^{\ast})}{2}$ and $A= [ N^{2/3}]+1$ in (\ref{asympsumkainf}), we see that 
	\[\sum_{k=[ N^{2/3}]+1}^{\infty} e^{-\frac{R'(x^{\ast})}{2N} \, k^2} \stackrel{N \rightarrow \infty}{\rightarrow} 0\;. \]
	
We proceed now to show that $\sum_{k=[ N^{2/3}]+1}^{J^{\ast}-1} e^{N(L(\frac{J^{\ast}-k}{N})-L(x^{\ast}))}$  tends to 0 when $N \rightarrow \infty$. In fact, in scenario $B\leftarrow \rightarrow A$ we know that $L''(x)=-\frac{R'(x)}{R(x)} <0$ for all $x \in[0,1]$. Letting $-M \equiv \max_{x\in[0,1]} L''(x)$, then $M>0$. Using a Taylor formula with Lagrange remainder, we know that there exists $\xi_k$ between $x^{\ast}$ and $\frac{J^{\ast}-k}{N}$ such that
\[N(L(\frac{J^{\ast}-k}{N})-L(x^{\ast})) \,=\, \frac{N}{2} \, L''(\xi_k) \left(\frac{J^{\ast}-k}{N}-x^{\ast}\right)^2 \, \leq -\frac{M}{2N} \,(k+\delta_N)^2 \;,\]
where we have used (\ref{yN}) and the above definition of $M$. As $\delta_N \geq -1/2$, we have
\[\sum_{k=[ N^{2/3}]+1}^{J^{\ast}-1} e^{N(L(\frac{J^{\ast}-k}{N})-L(x^{\ast}))} \,\leq \, \sum_{k=[ N^{2/3}]+1}^{J^{\ast}-1} e^{-\frac{M}{2N}(k-1)^2} \,\leq\, \sum_{k=[ N^{2/3}]}^{\infty} e^{-\frac{M}{2N}k^2}\;.\]
We may now use (\ref{asympsumkainf}) again and our claim follows.
$\blacksquare$

One last term remains to be controlled in order to complete the proof of Theorem \ref{theomainasymptotics}. We do so in
\begin{proposition} \label{properrorTaylor}
The term
\[\sum_{k=1}^{[ N^{2/3}]} \left[e^{N(L(\frac{J^{\ast}-k}{N})-L(x^{\ast}))}- e^{-\frac{R'(x^{\ast})}{2N} \, k^2}\right]\]
appearing in the proof of Theorem \ref{theomainasymptotics} is bounded when $N \rightarrow \infty$.
\end{proposition}

\textbf{Proof}
Notice first that $\frac{J^{\ast}-k}{N}= x^{\ast}- \frac{k+\delta_N}{N}$. Using the Taylor expansion (\ref{taylorexp3}), we may rewrite
\begin{eqnarray}
\sum_{k=1}^{[ N^{2/3}]} &&\left[e^{N(L(\frac{J^{\ast}-k}{N})-L(x^{\ast}))}- e^{-\frac{R'(x^{\ast})}{2N} \, k^2}\right]\,=\, \nonumber\\ &&\sum_{k=1}^{[ N^{2/3}]} e^{-\frac{R'(x^{\ast})}{2N} \, k^2} \left\{e^{-\frac{R'(x^{\ast})}{2N} [(k+\delta_N)^2-k^2] \,-\, \frac{L'''(x^{\ast}-\frac{\alpha_k (k+\delta_N)}{N})}{6N^2} \, (k+\delta_N)^3}-1 \right\} \;,\nonumber
\end{eqnarray}
where $\alpha_k$ is some number between 0 and 1.
We now use that for any $a>0$, 
\begin{equation}  \label{ethetabound}
|e^{\theta}-1| \leq \frac{e^a-1}{a} \,|\theta| \;,
\end{equation} 
if $|\theta| \leq a$. If we take 
\[\theta\,=\, -\frac{R'(x^{\ast})}{2N} [(k+\delta_N)^2-k^2] \,-\, \frac{L'''(x^{\ast}-\frac{\alpha_k (k+\delta_N)}{N})}{6N^2} \, (k+\delta_N)^3\]
and $M'=\max_{x \in[0,1]}|L'''(x)|$, then
\begin{eqnarray}
|\theta|&\leq & \frac{R'(x^{\ast})}{2N} \, |2\delta_N k+\delta_N^2| \,+\, \frac{M'}{6N^2} k^3 \, |1+\frac{\delta_N}{k}|^3 \nonumber\\
&\leq& \frac{R'(x^{\ast})}{8N} \,+\, \frac{R'(x^{\ast})}{2N}  \, k\,+\, \left(\frac{3}{2}\right)^3 \, \frac{M'}{6N^2} k^3 \nonumber \;,
\end{eqnarray}
where in the last passage we used that $|\delta_N|\leq 1/2$. As $k \leq N^{2/3}$ in the sum we want to estimate, there exists a constant $C$ independent of $N$ such that
$|\theta|<C$ if $k \leq N^{2/3}$.

By (\ref{ethetabound}), we obtain
that
\begin{eqnarray}
&&\left|e^{-\frac{R'(x^{\ast})}{2N} [(k+\delta_N)^2-k^2] \,+\, \frac{L'''(x^{\ast}-\frac{\alpha_k \delta_N}{N})}{6N^2} \, (k+\delta_N)^3}-1\right| \,\leq\, \frac{e^C-1}{C} \, |\theta| \nonumber\\
&\leq&  \frac{e^C-1}{C} \, \left(\frac{R'(x^{\ast})}{8N} \,+\, \frac{R'(x^{\ast})}{2N}  \, k\,+\,  \, \frac{9M'}{16N^2} k^3 \right) \;.\nonumber 
\end{eqnarray}
Substituting this bound and using (\ref{quadsum}) and (\ref{asympsumkp}), both with $p=1$ and $p=3$, finishes the proof.
$\blacksquare$

This last result appears in the proof of Theorem \ref{theocompletecoord}.
\begin{proposition} \label{propsmNx}
If the invasion scenario is $B\leftarrow \rightarrow A$ and $x<x^{\ast}$, then 
\[\sum_{k=1}^{Nx-1}R(x)^k\left[e^{N(L(x-\frac{k}{N})-L(x))-k \log R(x)}-1 \right] \stackrel{N \rightarrow \infty}{\rightarrow} 0\;.\]
\end{proposition}
\textbf{Proof}
By using for $L(x-\frac{k}{N})$ a Taylor expansion up to order 1 around $x$ with Lagrange remainder, then
\[N(L(x-\frac{k}{N})-L(x))-k \log R(x) \,=\, \frac{k^2}{2N}L''(x- \alpha_k \frac{k}{N})\]
for some $\alpha_k \in(0,1)$. If we use that $L''$ is negative in $[0,1]$ and for $\theta<0$ we have $|e^{\theta}-1|<|\theta|$, then
\[\left| \sum_{k=1}^{Nx-1}R(x)^k\left[e^{N(L(x-\frac{k}{N})-L(x))-k \log R(x)}-1 \right]\right| \,<\, \frac{M}{2N}\, \sum_{k=1}^{Nx-1}k^2 R(x)^k \;,\]
where $M =\max_{y \in [0,1]}|L''(y)|$. As the latter sum converges if we exchange the upper limit by $\infty$, the proposition is proved.
$\blacksquare$

\section*{Acknowledgements}
We thank Max O. Souza for early discussions and encouragement for writing this paper.

\end{document}